\documentclass[fleqn,usenatbib]{mnras}

\usepackage{newtxtext,newtxmath}

\usepackage[T1]{fontenc}

\DeclareRobustCommand{\VAN}[3]{#2}
\let\VANthebibliography\thebibliography
\def\thebibliography{\DeclareRobustCommand{\VAN}[3]{##3}\VANthebibliography}

\usepackage{graphicx}	
\usepackage{amsmath}	
\usepackage{caption}
\usepackage{longtable}
\usepackage{multirow}
\usepackage{pdflscape}
\usepackage{booktabs}

\title[Diffuse interstellar bands in the near-IR]{Diffuse interstellar bands in the near-infrared: Expanding the reddening range}

\author[R. Castellanos et al.]{
R.Castellanos,$^{1,2}$\thanks{E-mail: rcastellanos@cab.inta-csic.es}
F. Najarro,$^{1}$
M. Garcia,$^{1}$
L. R. Patrick,$^{1}$
T. R. Geballe$^{3}$
\\
$^{1}$Departamento de Astrofísica, Centro de Astrobiología (CSIC-INTA), Ctra. Torrejón a Ajalvir km 4, E-28850 Torrejón de Ardoz, Spain\\
$^{2}$Departamento de Física Teórica, Universidad Autónoma de Madrid (UAM), Campus de Cantoblanco, E-28049 Madrid, Spain\\
$^{3}$Gemini Observatory/NSF's NOIRLab, 670 N. Aohoku Pl., Hilo, HI 96720, USA
}

\date{Accepted XXX. Received YYY; in original form ZZZ}

\pubyear{2024}

\begin{document}
\label{firstpage}
\pagerange{\pageref{firstpage}--\pageref{lastpage}}
\maketitle

\begin{abstract}
We have investigated the behaviour of three strong near-infrared diffuse interstellar bands (DIBs) at $\lambda$13177\,Å, $\lambda$14680\,Å, and $\lambda$15272\,Å, on a larger sample of sightlines and over a wider range of extinctions than previously studied, utilizing spectra from three observatories. 
We applied two telluric correction techniques to reduce atmospheric contamination and have used Gaussian fits to characterise the DIB profiles and measure equivalent widths.
We confirmed strong and approximately linear correlations with reddening of the $\lambda$13177\,Å, $\lambda$14680\,Å and $\lambda$15272\,Å DIBs, extending them to higher reddening values and strengthening their link to interstellar matter. 
Modelling of the $\lambda$14680\,Å DIB profiles revealed intrinsic variations, including line broadening, linked to their formation processes. This effect is particularly pronounced in the Galactic Centre (GC) environment, where multiple diffuse molecular clouds along the line of sight contribute to line broadening.
We have detected one new DIB candidate at $\lambda$14795\,Å on sightlines with high reddening.
\end{abstract}

\begin{keywords}
ISM: clouds – ISM: lines and bands – ISM: molecules – dust, extinction – stars: early-type 
\end{keywords}



\section{Introduction}

The diffuse interstellar bands (DIBs) comprise more than 500 absorption features detectable on sightlines towards sources obscured by interstellar clouds \citep{hobbs2008catalog, hobbs2009studies}.
The term "diffuse" refers to their broad profiles compared to profiles of atomic or molecular lines. 
Measuring their intensities is a complex task since determining equivalent widths (EWs) depends on accurate placements of stellar continua and on the removal of stellar lines. 
Hot, early-type stars, which contain few and often weak spectral lines, are the best background illuminators of these bands.

The interstellar origin of DIBs was established by \citet{merrill1938unidentified} by correlating their intensities with reddening. 
However, despite being first detected 100 years ago, the identification and production processes of the DIBs, except in a few cases, remain unexplained.
Initially, it was believed that dust particles were responsible for the DIBs due to the correlation of DIBs EW with colour excess.
However, searches for polarization in DIB profiles indicate minimal polarization efficiency of the carriers, suggesting that they are in the gas phase rather than being attached to dust grains \citep{cox2011linear}. 
Although the DIBs are not located on dust particles, dust may still play a crucial role in protecting the actual DIB molecular carriers from ultraviolet radiation. 
Nevertheless, DIBs may appear toward slightly reddened objects where their carriers can withstand the local ultraviolet radiation \citep{galazutdinov1998diffuse}.

Complex spectral substructures revealed in the narrowest DIBs through high-resolution spectroscopy \citep{sarre1995resolution, ehrenfreund1996resolved, galazutdinov2008high} suggest that complex carbon molecules, in particular Polycyclic Aromatic Hydrocarbons (PAHs), are the principal carriers of these absorption features. 
However, the identification of the carriers using gas-phase laboratory analysis is challenging, as there are over a million possible PAH species that contain less than 100 carbon atoms that could serve as carriers \citep{cami2018eso}.
Investigating the behaviour of DIBs across different lines of sight, especially the particular targets affected by a single interstellar cloud, is critical to identifying their carriers and understanding how the physical properties of interstellar clouds impact DIB profile variations \citep{elyajouri2017near}.

The DIBs can be categorised into two main groups according to their widths. 
The first and larger group are those DIBs with narrow profiles, which extends approximately over 1\,Å and typically displays substructure. 
The second group is comprised of "diffuse" absorption profiles, having widths exceeding 10\,Å and usually exhibiting non-Gaussian profiles \citep{krelowski2018diffuse}. 
Depending on the shape of the profile, one can distinguish two different types of clouds in which the DIB arises \citep{galazutdinov2008high}.

While past research on DIBs has predominantly focused on the optical range \citep{raimond2012southern, cox2017eso}, it is noteworthy that charged PAHs containing over 20 carbon atoms display their main electronic transitions in the near infrared (NIR) \citep{mattioda2005experimental, ruiterkamp2005pah, tielens2008interstellar}. 
Consequently, if some bands are associated with PAHs one would expect to find DIBs in the NIR.
However, infrared studies of DIBs such as the one conducted by \citet{Hamano_2015} have until fairly recently been less frequent due to the later arrival of sensitive high-resolution spectrographs in the NIR and to difficulties in accurately correcting the spectra for telluric absorption lines.

Two distinct features at $\lambda$11795\,Å and $\lambda$13177\,Å (wavelengths in vacuo) discovered by \citet{joblin1990detection} were the first DIBs detected at wavelengths greater than one micron. 
The high-resolution profiles (R ($\lambda$ / $\delta\lambda$) $\sim$ 37,000) of these DIBs observed by \citet{rawlings2014high} showed that they had the typical asymmetry characterised by an enhancement of the red wing, seen in most broad diffuse bands. 
\citet{geballe2011infrared} discovered 13 new diffuse bands in the 1.5-1.8 \textmu m interval, including the prominent $\lambda$15272\,Å  DIB. 
\citet{cox2014vlt} validated nine of these bands and discovered 11 new NIR DIBs as well.
\citet{Hamano_2015} identified fifteen DIBs in the 9100-13200 Å range including seven of the DIBs proposed as candidates by \citet{cox2014vlt}.
Subsequently, \citet{Hamano_2016} reported the existence of a family of DIBs in the NIR spectral region, which included the DIBs of $\lambda$10780, $\lambda$10792, $\lambda$11797, $\lambda$12623, and $\lambda$13175 Å.

In this paper, we aim to improve our understanding of several of these longer wavelength DIBs by observing them on a large sample of sightlines covering a wide range of extinctions.
While significant progress has been made in characterizing the behaviours of the $\lambda$13177\,Å \citep{Hamano_2015, smoker2022high} and $\lambda$15272\,Å \citep{Elyajouri_L_2017, smoker2022high} DIBs, the presence and properties of the other DIBs at wavelengths beyond one micron remain relatively unexplored. 
One focus is on comparing the properties of those two DIBs with the DIB at $\lambda$14680\,Å, which prior to this work has only been reported by \citet{galazutdinov2017infrared} in two sources.
By focusing our research on the near infrared bands, we are able to include sources with large reddening (Av > 10) which are too faint to be studied at optical wavelengths. 

The value of studying infrared DIBs is enhanced by recent launches into space of IR observing facilities. 
To fully exploit the potential of modern and future space observatories, it is necessary to explore all potential sources of spectral features, including DIBs. 
Our study is timely as the $\lambda$14680\,Å DIB and a newly discovered DIB at $\lambda$14795\,Å are at wavelengths of poor atmospheric transmission between the J and H bands which are fully accessible from space.

This paper is organized as follows: in Sect. \ref{sec:Observations} we present the observations and data reduction for a selection of sightlines through the diffuse interstellar medium to bright early-type stars. 
The observations were conducted using four spectrometers at three observatories. In Sect. \ref{sec:Results} we present spectra, EWs and full-widths-at-half-maximum (FWHMs) of the NIR DIBs that were observed. 
In Sect. \ref{sec:Discussion} we discuss the properties of the $\lambda$13177\,Å, $\lambda$14680\,Å and $\lambda$15272\,Å DIBs in particular, their strengths and profiles shape together with their correlations with extinction and with each other. 
The main conclusions of this survey are summarised in Sect. \ref{sec:Conclusions}.

\section{Observations and Data Acquisition}
\label{sec:Observations}

\begin{figure}
	\includegraphics[width=\columnwidth]{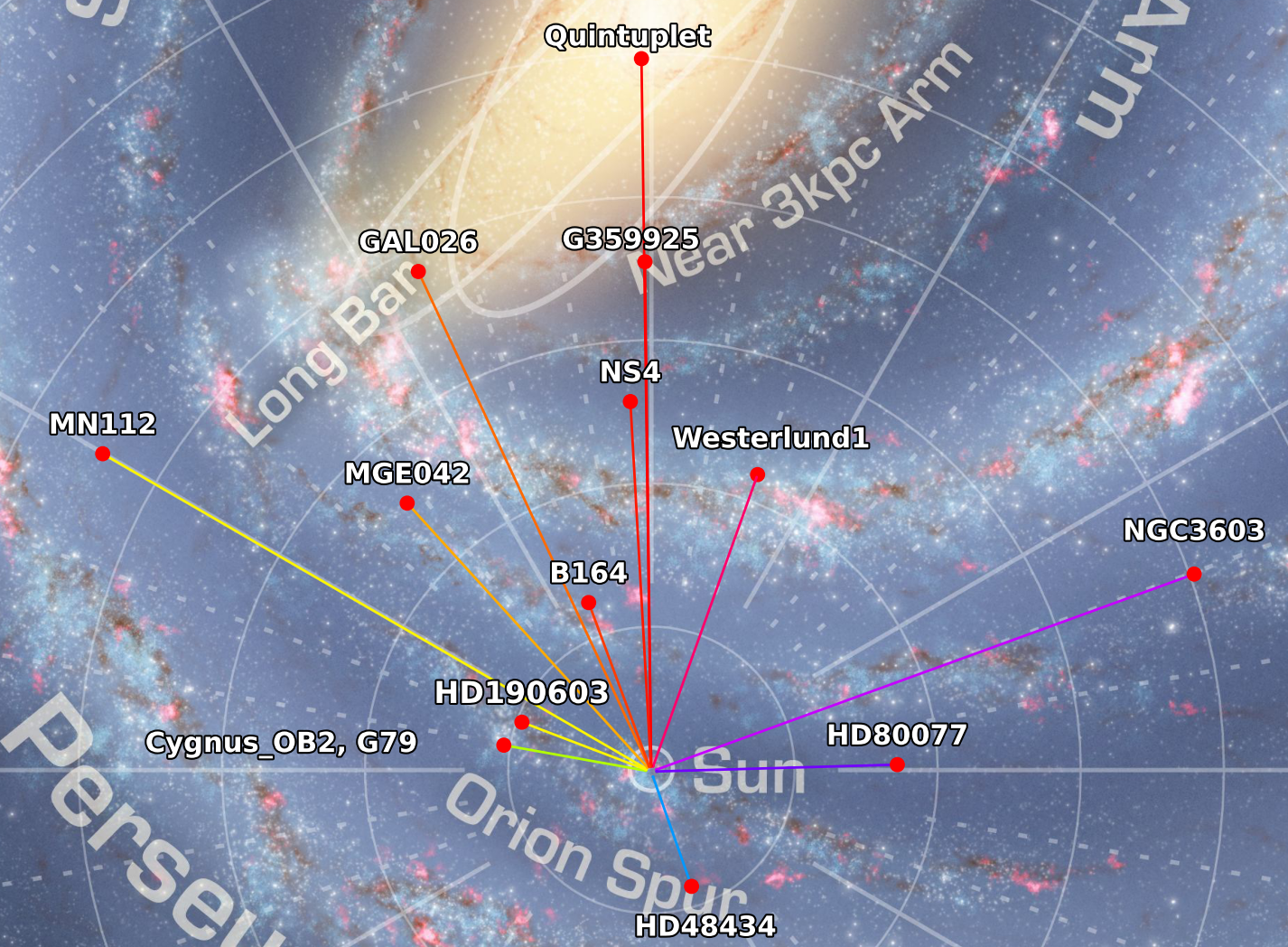}
    \caption{Galactic map showing the lines of sight compiled for our study. Each line represents the direction and position of a surveyed object, colour-coded consistently with the figures in the following sections. Adjacent concentric circles are separated by 5000 light years ($\sim$ 1.53 kpc)}.
    \label{fig:mapa_galaxia}
\end{figure}

This study focuses on the wavelength interval 13000-15300\,Å and utilises spectroscopic observations obtained from a diverse set of stars distributed across the Galactic plane.
Their spatial distribution is illustrated in Fig. \ref{fig:mapa_galaxia}.
In this study, we utilized the $E(J-K)$ colour index as an approximation of reddening values. \footnote{According to the extinction law by \citet{Cardelli_1989}, the range of $E(J-K)$ values in our sample corresponds to an approximate $A_V$ range of 0 to 30 magnitudes.}
This method is justified as our targets, predominantly massive stars, exhibit minimal intrinsic $E(J-K)$ value variations, typically up to 0.15. 
This variation is negligible, especially compared to the range of $E(J-K)$ values being examined, making it a suitable proxy for reddening.
The study covers a broad range of reddening values, spanning the range from stars with virtually no reddening used as anchoring points to stars with large reddening in the stellar clusters situated at the GC. 
A summary of the observations is given in Table~\ref{tab:Observations}.
A more detailed list of objects is given in Table~\ref{tab:DIBs_Values}.
Additional information on the distance and location of each object is provided in Table~\ref{tab:Targets_info}.

\begin{table}
\centering
\caption{Summary of observations.}
\label{tab:Observations}
\begin{tabular}{ccccc}
\toprule
Nº of objects & Band & Telescope/Instrument & R($\lambda$/$\Delta$$\lambda$) \\
\midrule
6 & H & VLT/KMOS & 4000 \\
16 & J, H & VLT/KMOS & 3600, 4000 \\
10 & J, H & GEMINI/GNIRS & 5900 \\
13 & J, H, K & VLT/Xshooter & 5600-11600 \\
11 & J, H, K & TNG/GIANO-B & 50000 \\
\bottomrule
\end{tabular}
\end{table}

\subsection{Telluric correction}
\label{sec:Telluric_correction} 

In the 13000-15300\,Å interval, the transmission spectrum of the Earth's atmosphere contains a multitude of strong absorption lines of water vapour which significantly contaminate spectra of astronomical objects.
Some of the data analyzed in the paper were already fully reduced and published.
For the unreduced spectra, we have adopted two different approaches to telluric correction, tailored to the specific instruments used in our study.

Our initial approach for telluric correction employs the traditional and usually effective strategy that involves the observation of telluric standard stars at approximately equivalent airmasses as the primary scientific target observations taken immediately before or after.
These stars, typically early-type dwarfs, preferentially A0V with relatively featureless NIR spectra, serve as atmospheric references. 
We analysed the spectra of telluric standard stars to identify and model their intrinsic stellar features and subtract them to create a pure telluric transmission spectrum.
The fully reduced spectrum of the target star is obtained by dividing the spectrum of the object by the telluric spectrum.

Our second method relies on utilizing atmospheric transmission models and libraries to accurately remove atmospheric absorption features from observed spectra. 
We have used the Molecfit tool \citep{smette2015molecfit} which is integrated into several VLT data reduction pipelines. 
Once properly tuned, Molecfit has proven to be very efficient on the VLT-KMOS and VLT-Xshooter datasets. 
An example of the precision that can be obtained is shown in Fig. \ref{fig:Telluric_correction}.

The choice of the telluric correction methodology relied on the data processing tools available for each instrument. 
When both the Molecfit software and the standard star method were available, we
conducted a comparative analysis to determine the most effective method. 
We focused on achieving the lowest residuals specifically in the spectral regions where the DIBs are located.
In the following subsections, we describe the observations.

\subsection{VLT-KMOS}
\label{sec:KMOS} 

The K-band Multi-Object Spectrograph (VLT-KMOS; \citealt{Sharples_2013}) was utilized to gather data on the massive stellar clusters NGC3603 and Westerlund1 (Wd1) with R($\lambda$/$\Delta$$\lambda$) values of approximately 3600 and 4000 for the J and H bands, respectively. 
These observations were conducted as part of two ESO programs: 098.D-0373(A) (P.I.: C. Evans) contributed two sources from the NGC3603 cluster, while program 109.233D.001 (P.I.: R. Castellanos) provided data for 14 sources from the Wd1 cluster. 

Seven additional sources were included from ESO programme 0103.D-0316(A) (P.I.: S. Clark), which involve spectroscopy of massive stars within the Quintuplet cluster. 
This programme only covered the H and K bands. 
Consequently, there is no available data for the $\lambda$13177\,Å DIB for these sources, as indicated in Table \ref{tab:DIBs_Values}.

To perform the data reduction of those programmes mentioned above that have not yet been published, we used the KMOS pipeline (SPARK, \citealt{Davies_2013}) in the ESO Reflex automated data reduction environment \citep{Freudling_2013}.
To effectively address telluric effects in the KMOS dataset, we employed the Molecfit software package but also tested the telluric standard approach (Castellanos et al. in prep) which yielded consistent results.

\subsection{VLT-Xshooter}
\label{sec:Xshooter} 

We compiled VLT-Xshooter \citep{Vernet_2011} archive data from ESO Phase 3, covering 13 individual targets with different reddening values. 
The spectral resolution of the spectra gathered from VLT-Xshooter archive varies between R $\sim$ 5600 - 11600, depending on the slit width used for the observations.
Since no telluric standards were available for most of the sources, we used Molecfit to perform the telluric correction for this dataset.

\subsection{TNG-GIANO-B}
\label{sec:GIANO-B} 

TNG-GIANO-B \citep{oliva2012a,oliva2013giano} high-resolution spectra  (R $\approx$ 50,000) of nine sources from program CAT19A\_124 (P.I.: F. Najarro) were compiled. 
These sources predominantly consist of bright O and B stars, along with luminous blue variables (LBVs). 
For this data reduction, we utilized the automated GOFIO software \citep{Rainer_2018}, specifically designed for handling the data produced by the TNG-GIANO-B instrument.
To remove the telluric lines from the TNG-GIANO-B data, we used the spectra of standard telluric stars as described in Sect.\ref{sec:Telluric_correction}.
Additionally, we used Molecfit on the data (standalone version 2.0) and obtained similar results.

\subsection{GEMINI-GNIRS}
\label{sec:GEMINI} 

We analyzed the DIBs in fully reduced and in some cases previously published spectra \citep{geballe2011infrared} obtained at the Gemini North Telescope with the facility GNIRS \citep{Elias_2006}.
The spectra are mainly of highly reddened massive stars in the GC. 
Similar to TNG-GIANO-B, spectra of standard stars observed on the same night as the program stars were used to perform the telluric correction.

\section{Results}
\label{sec:Results}

\begin{figure}
	\includegraphics[width=\columnwidth]{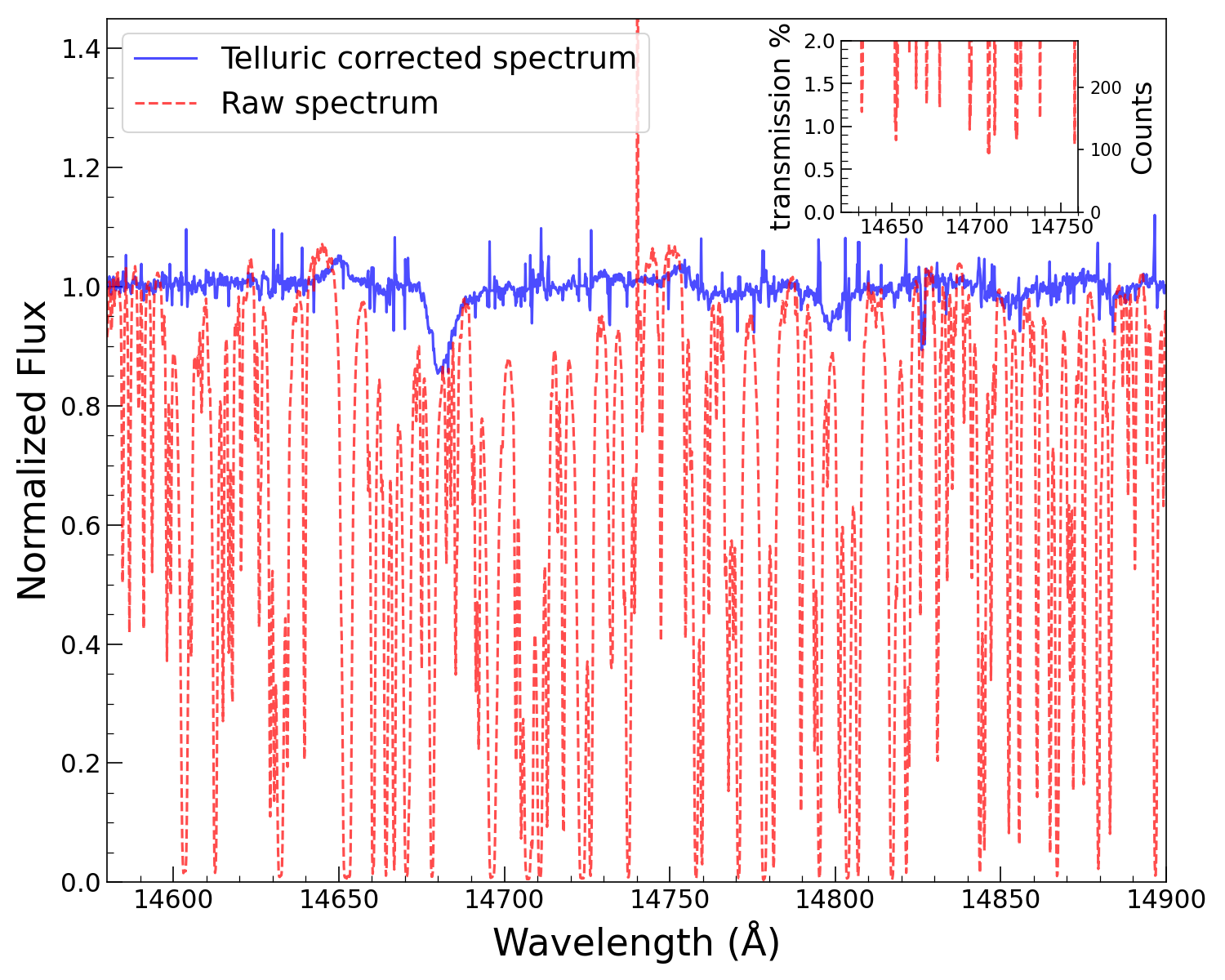}
    \caption{TNG-GIANO-B observations of TIC 109609771 LBV in the 14550-14900\,Å wavelength range, heavily impacted by telluric contamination. The blue line represents the telluric-corrected spectrum, revealing the presence of the $\lambda$14680\,Å DIB and the newly discovered DIB at $\lambda$14795\,Å, while the red dashed line depicts the raw spectrum. Inset (top-right) shows that transmission does not fall below 0.75\% ($\sim$100 counts) in the region around this DIB and therefore a clean telluric removal can be carried out provided we have a large S/N}.
    \label{fig:Telluric_correction}
\end{figure}

\subsection{Detection of NIR DIBs}

\begin{figure*}
	\includegraphics[width=\textwidth]{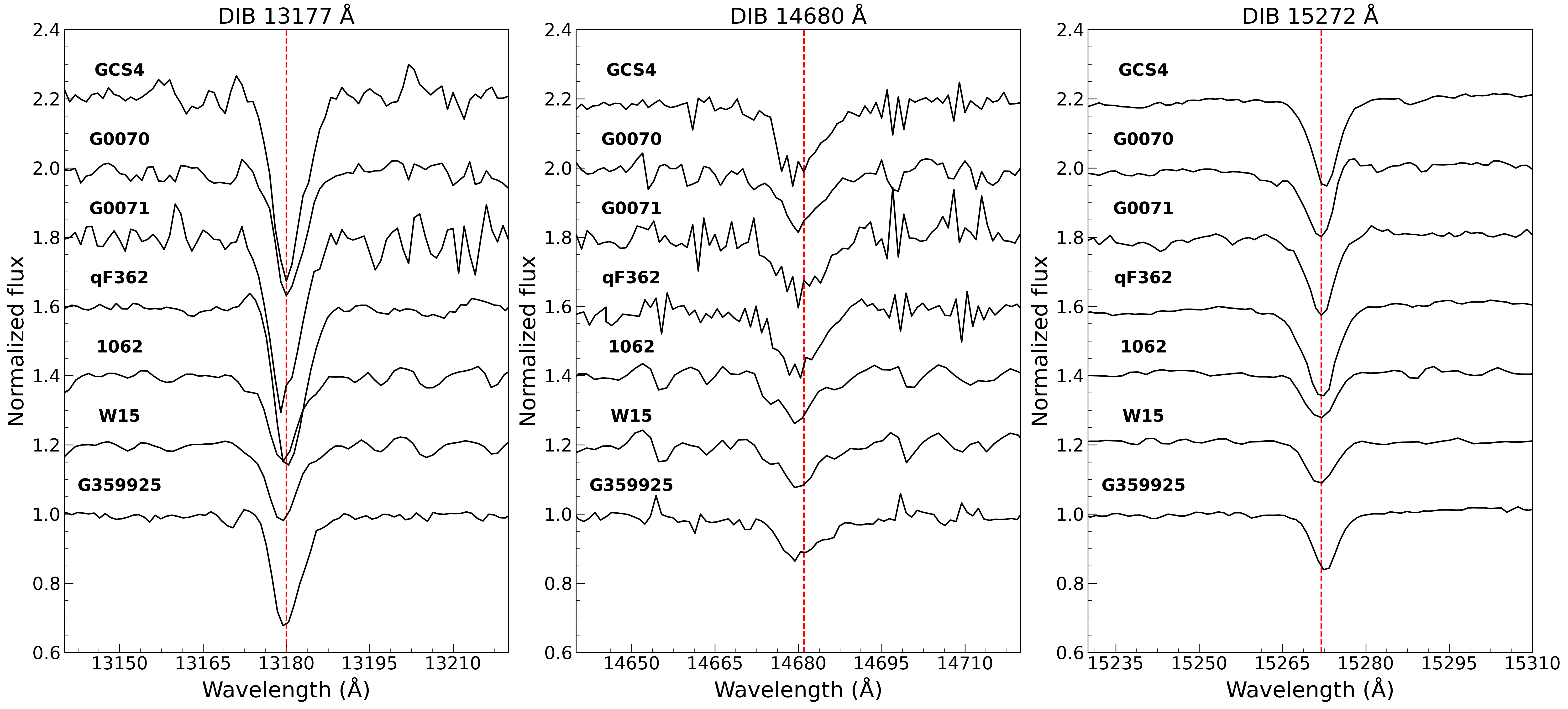}
    \caption{NIR DIBs at $\lambda$13177\,Å (left panel), $\lambda$14680\,Å (middle panel), and $\lambda$15272\,Å (right panel) are shown for a sample of sightlines. The spectra have been corrected for the presence of telluric lines and corrected to the Local Standard Rest. The dashed lines indicate the rest wavelength for the respective DIBs.}
    \label{fig:Examples_DIBs}
\end{figure*}

Fig. \ref{fig:Examples_DIBs} shows a selection of reduced and normalized spectra.
The strong DIBs, located at $\lambda$13177\,Å and $\lambda$15272\,Å, are detected towards all stars having a $E(J-K)$ indices larger than 0.1. 
We also confirm the presence of a strong DIB at $\lambda$14680\,Å which had escaped detection in previous studies in the literature, except for its presence in two targets from \citet{galazutdinov2017infrared}.
This DIB is located in the telluric line forest at the short wavelength edge of the H-band (see Fig.\ref{fig:Telluric_correction}).
In order to detect and characterize it, stable and relatively dry atmospheric conditions and precise telluric removal are essential.

\begin{figure}
	\includegraphics[width=\columnwidth]{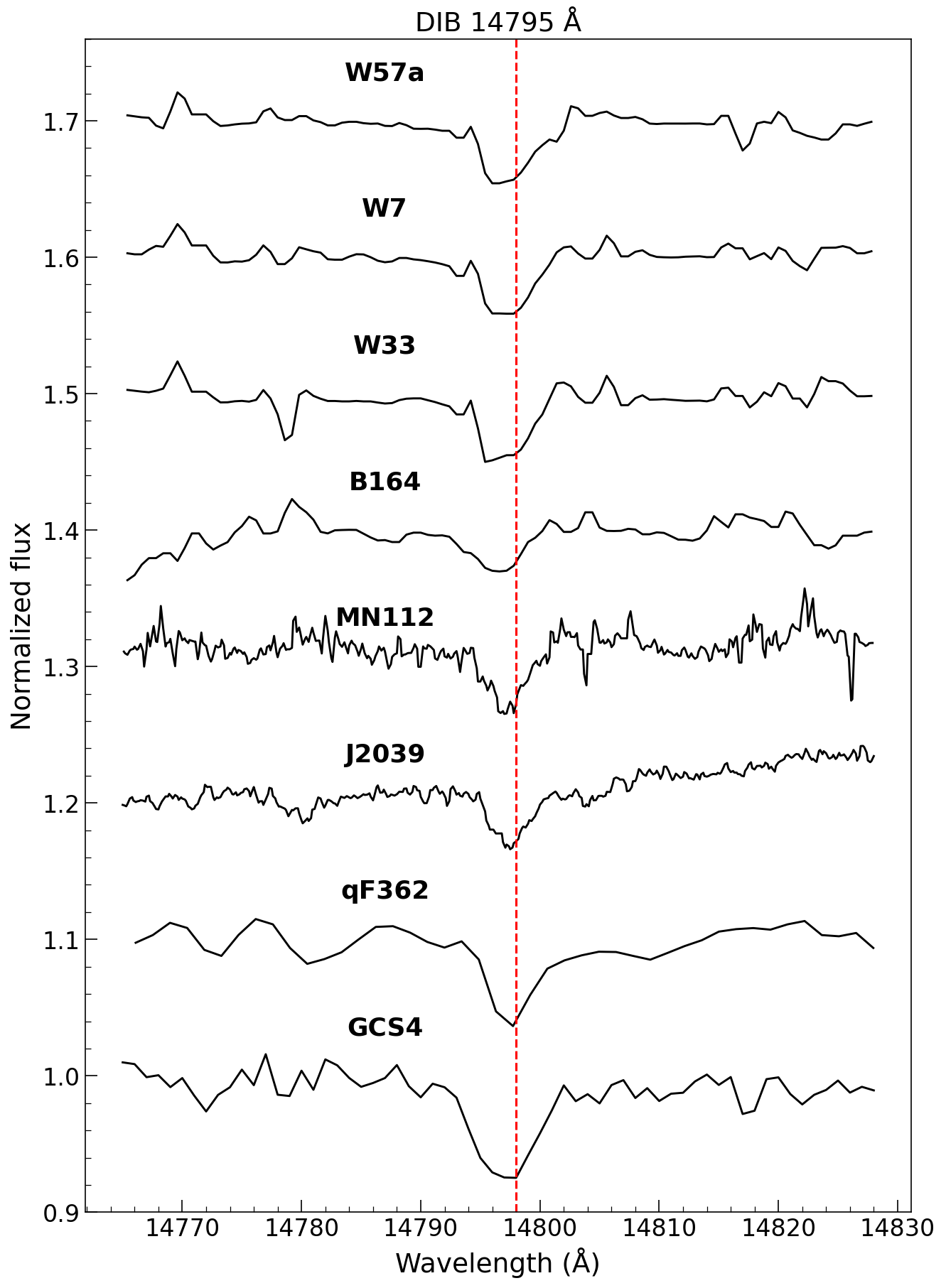}
    \caption{Spectra of eight representative sources in the vicinity of the newly detected $\lambda$14795\,Å DIB. The dashed line indicate the rest wavelength for the DIB.}
    \label{fig:Examples_DIB_14795}
\end{figure}

\begin{figure}
	\includegraphics[width=\columnwidth]{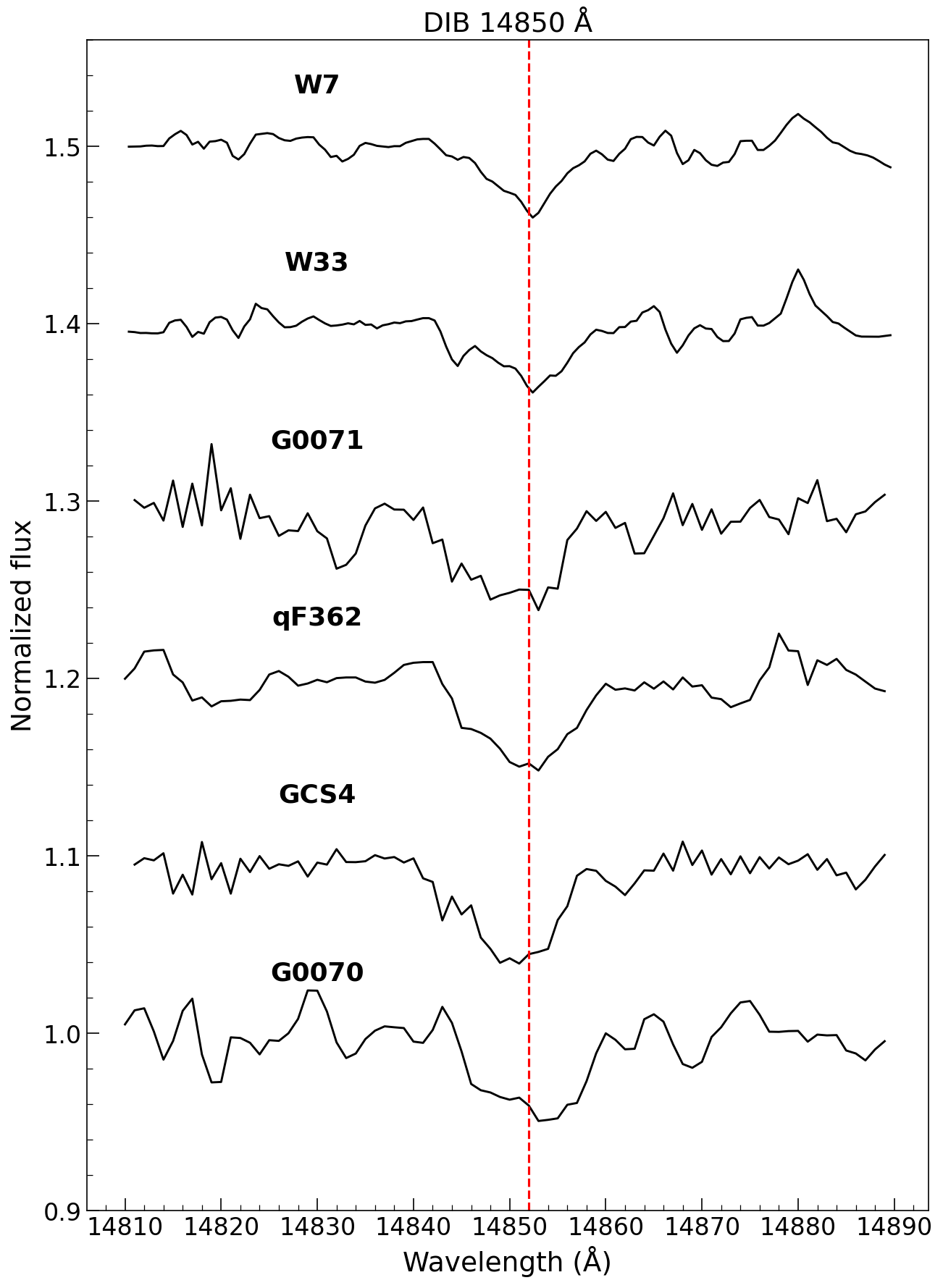}
    \caption{Spectra of six representative sources in the vicinity of the $\lambda$14850\,Å DIB. The dashed line indicate the rest wavelength for the DIB.}
    \label{fig:Examples_DIB_14850}
\end{figure}

Additionally, we identify a new DIB at $\lambda$14795\,Å, detecting it and the nearby $\lambda$14850\,Å DIB \citep{Najarro_GC_2017} in a small number of sources with significant reddening (see Figs. \ref{fig:Examples_DIB_14795} and \ref{fig:Examples_DIB_14850}).
The detectability of these absorptions is severely constrained by the necessity for a high S/N on the stellar continuum due to their weakness.  
Despite this, the detection of these two DIB candidates in several sources, together with their common behaviour and consistent profile, strengthens their identification as DIBs. 
The $\lambda$14795\,Å DIB exhibits a narrow profile, as illustrated in Fig. \ref{fig:Examples_DIB_14795}, whereas the $\lambda$14850\,Å band displays a broader profile (see Fig. \ref{fig:Examples_DIB_14850}; and Sect.\ref{sec:New_DIBs} for more details).  

We summarise the NIR DIBs reported in our sample as follows:
\begin{itemize}
   \item The DIB at $\lambda$13177\,Å is detected in 45 sources in the J band.
   \item The DIB at $\lambda$14680\,Å is found in 51 sources in the H band.
   \item The DIB at $\lambda$15272\,Å is observed in 51 sources in the H band.
   \item The newly identified $\lambda$14795\,Å DIB is detected in 14 sources in the H band.
   \item The $\lambda$14850\,Å DIB is detected in 7 sources in the H band.
\end{itemize}

\subsection{DIB strength and fitting method}

In this section we describe our method of deriving the DIB properties.
First, we normalize the segment of the spectrum near the DIB by fitting the continuum on either side of the band with polynomials of order three or four and dividing the observed spectrum by that continuum fit.
We consider as detected, and therefore included in our sample, only those DIBs for which the measured EW exceeds five times the standard deviation calculated from adjacent regions on either side of the band. 
For these DIBs we fit the observed profile with either a single Gaussian or the superposition of two Gaussians.

\begin{figure*}
	\includegraphics[width=\textwidth]{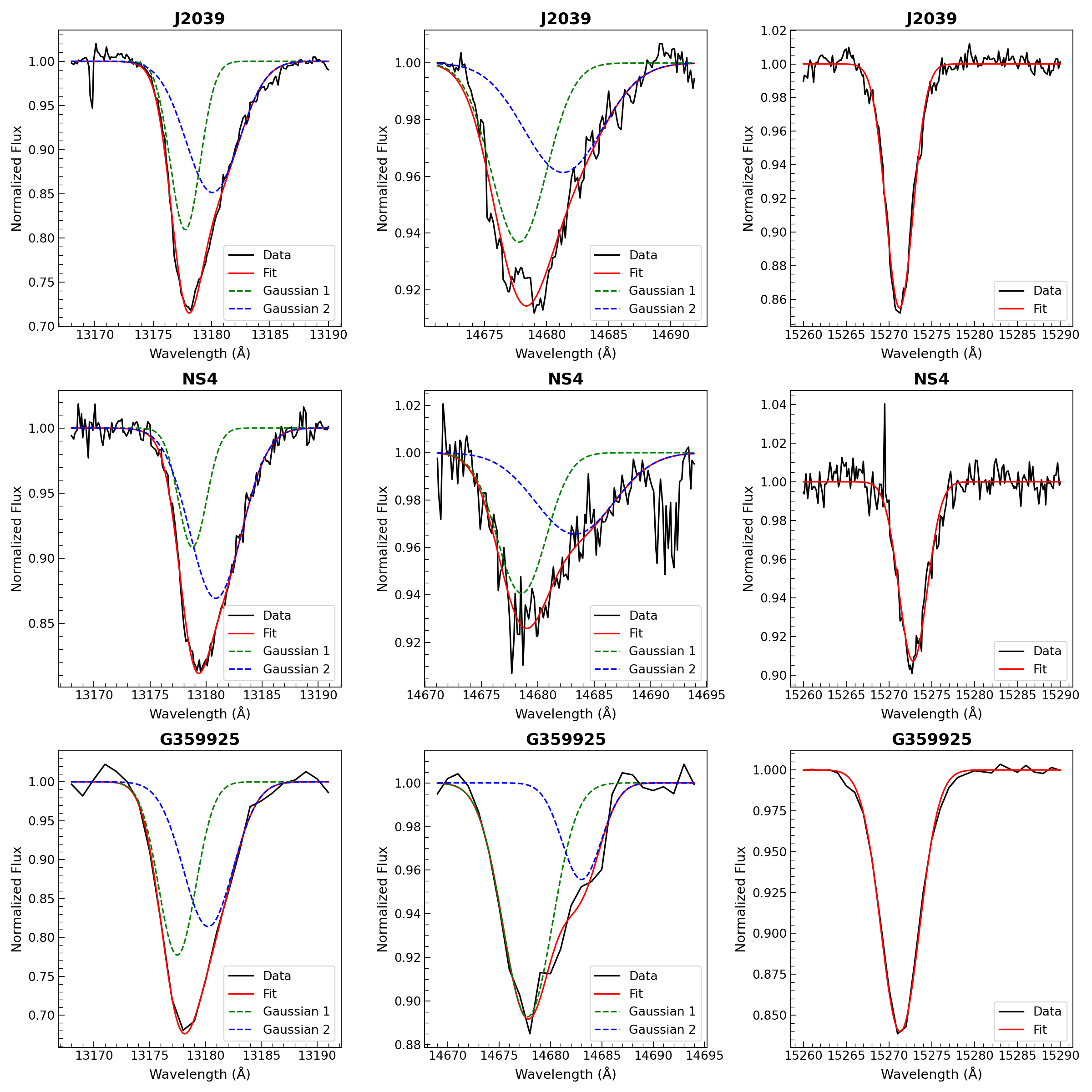}
    \caption{Normalized spectra for three selected stars: J2039, NS4, and G359925 in DIBs at $\lambda$13177\,Å (left), $\lambda$14680\,Å (centre), and $\lambda$15272\,Å (right). Observed spectra: solid black lines; final fit: red lines; Gaussian components: dashed lines.}
    \label{fig:Gaussian_fits}
\end{figure*}

Fig. \ref{fig:Gaussian_fits} presents the spectra of a few sources that were observed using TNG-GIANO-B (upper and mid panels) and GEMINI-GNIRS (lower panel), and the single Gaussian or two-Gaussian fits, applied to the $\lambda$13177\,Å, $\lambda$14680\,Å, and $\lambda$15272\,Å DIBs profiles.
The $\lambda$13177\,Å and $\lambda$14680\,Å DIBs have asymmetric profiles, with long wavelength shoulders, and are better fit by two Gaussians than by a single Gaussian.

The results of the Gaussian fitting are summarised in Table \ref{tab:DIBs_Values}.
The EWs calculated from the fits are consistent with those obtained from numerical integrations over the line profiles.
We adopt the methodology recommended by \citet{bodensteiner2020young} to assess the uncertainties associated with the EW measurement following \citet{Chalabaev_1983}:

\[
\sigma_{\text{EW}} = \frac{\sqrt{2 \cdot w \cdot \Delta\lambda}}{S/N}
\]

where S/N is the signal-to-noise ratio of the continuum adjacent to band and $\Delta\lambda$ is the wavelength step and \textit{w} represents the width of the window used to measure the EW, which varies for each DIB depending on its intensity.
Systematic errors, mainly linked to continuum normalisation, could lead to larger uncertainties.

\clearpage
\onecolumn
\begin{landscape}
\footnotesize
\setlength\tabcolsep{4pt} 
\setlength{\LTpost}{0pt} 
\begin{longtable}{lcccccccccccccccc}
\caption{List of objects, EWs and FWHM values for three NIR DIBs.} \label{tab:DIBs_Values} \\
\hline 
\multirow{2}{*}[-1em]{Star \textit{(Alt. ID)}} & \multirow{2}{*}[-1em]{Instrum} & \multirow{2}{*}[-1em]{$\underset{\text{(º)}}{\text{Sightline\footnotemark[1]}}$} & \multirow{2}{*}[-1em]{Sp type} & \multirow{2}{*}[-1em]{J} & \multirow{2}{*}[-1em]{K} & \multirow{2}{*}[-1em]{Ref.\footnotemark[2]} & \multirow{2}{*}[-1em]{$E(J-K)$\footnotemark[3]} & \multicolumn{3}{c}{\raisebox{-1.5ex}{$\lambda$13177\,Å}} & \multicolumn{3}{c}{\raisebox{-1.5ex}{$\lambda$14680\,Å}} & \multicolumn{3}{c}{\raisebox{-1.5ex}{$\lambda$15272\,Å}} \\
& & & & & & & & & & & & & & & & \\
\cline{9-11}\cline{12-14}\cline{15-17}
& & & & & & & & S/N & $\underset{\text{(Å)}}{\text{$W_{\lambda}$}}$ & $\underset{\text{(Å)}}{\text{FWHM}}$ & S/N & $\underset{\text{(Å)}}{\text{$W_{\lambda}$}}$ & $\underset{\text{(Å)}}{\text{FWHM}}$ & S/N & $\underset{\text{(Å)}}{\text{$W_{\lambda}$}}$ & $\underset{\text{(Å)}}{\text{FWHM}}$ \\
\hline
\hline
\endfirsthead

\multicolumn{17}{c}%
{{\tablename\ \thetable{} -- \textit{continued.}}} \\
\hline
\multirow{2}{*}[-1em]{Star \textit{(Alt. ID)}} & \multirow{2}{*}[-1em]{Instrum} & \multirow{2}{*}[-1em]{$\underset{\text{(º)}}{\text{Sightline\footnotemark[1]}}$} & \multirow{2}{*}[-1em]{Sp type} & \multirow{2}{*}[-1em]{J} & \multirow{2}{*}[-1em]{K} & \multirow{2}{*}[-1em]{Ref.\footnotemark[2]} & \multirow{2}{*}[-1em]{$E(J-K)$\footnotemark[3]} & \multicolumn{3}{c}{\raisebox{-1.5ex}{$\lambda$13177\,Å}} & \multicolumn{3}{c}{\raisebox{-1.5ex}{$\lambda$14680\,Å}} & \multicolumn{3}{c}{\raisebox{-1.5ex}{$\lambda$15272\,Å}} \\
& & & & & & & & & & & & & & & & \\
\cline{9-11}\cline{12-14}\cline{15-17}
& & & & & & & & S/N & $\underset{\text{(Å)}}{\text{$W_{\lambda}$}}$ & $\underset{\text{(Å)}}{\text{FWHM}}$ & S/N & $\underset{\text{(Å)}}{\text{$W_{\lambda}$}}$ & $\underset{\text{(Å)}}{\text{FWHM}}$ & S/N & $\underset{\text{(Å)}}{\text{$W_{\lambda}$}}$ & $\underset{\text{(Å)}}{\text{FWHM}}$ \\
\hline
\hline
\endhead

\hline
\multicolumn{17}{r}{{c}} \\
\endfoot

\hline
\vspace{-20mm} 
\endlastfoot

    WR42e & KMOS & 291.57 & O2If*/WN6 & 10.18 & 9.04 & 1 & 1.35 & 90 & 1.01 $\pm$ 0.06 & 6.66 &  82 & 0.62 $\pm$ 0.11 & 7.66  &  200 & 0.49 $\pm$ 0.04 & 6.54 \\
    TIC 467481534 (\textit{MTT31})& KMOS & 291.62 & O2V & 10.67 & 9.86 & 1 & 1.02 & 50 & 0.97 $\pm$ 0.13 & 5.70 & 113 & 0.67 $\pm$ 0.09 & 8.12 & 200 & 0.54 $\pm$ 0.04 & 6.03 \\
    \lbrack CRN2020\rbrack\ W1008 & KMOS & 339.52 & O9.5 II & 11.93 & 10.41 & 3 & 1.66 & 67 & 1.43 $\pm$ 0.12 & 6.92 &  67 & 1.09 $\pm$ 0.13 & 11.02 & 142 & 0.75 $\pm$ 0.05 & 5.99 \\
    \lbrack CRN2020\rbrack\ W1025 & KMOS & 339.53 & O+O? & 12.98 & 11.48 & 3 & 1.68 & 144 & 1.46 $\pm$ 0.07 & 6.03 & 63 & 1.15 $\pm$ 0.14 & 8.03 & 218 & 0.88 $\pm$ 0.08 & 7.27 \\
    \lbrack CRN2020\rbrack\ W1034 & KMOS & 339.53 & O9.5 Iab & 11.49 & 10.0 & 3 & 1.63 & 100 & 1.41 $\pm$ 0.09 & 6.55 & 69 & 1.19 $\pm$ 0.13 & 10.02 & 90 & 0.91 $\pm$ 0.07 & 6.64 \\
    \lbrack CRN2020\rbrack\ W1043 & KMOS & 339.55 & O9.5 II-III & 11.78 & 10.09 & 3 & 1.83 & 191 & 1.52 $\pm$ 0.05 & 6.78 & 83 & 1.12 $\pm$ 0.11 & 11.31 & 246 & 0.78 $\pm$ 0.08 & 6.84 \\
    \lbrack CRN2020\rbrack\ W1047 & KMOS & 339.53 & O9.5 II & 11.53 & 9.98 & 3 & 1.69 & 107 & 1.48 $\pm$ 0.09 & 6.52 &  65 & 1.11 $\pm$ 0.16 & 7.45 & 180 & 0.8 $\pm$ 0.05 & 6.06 \\
    \lbrack CRN2020\rbrack\ W1062 & KMOS & 339.56 & O+O? & 12.44 & 10.85 & 3 & 1.77 & 97 & 1.52 $\pm$ 0.07 & 5.87 & 52 & 1.17 $\pm$ 0.16 & 8.96 & 138 & 0.78 $\pm$ 0.04 & 6.15 \\
    \lbrack CRN2020\rbrack\ W1065 & KMOS & 339.58 & B0 Ib & 10.06 & 8.25 & 3 & 1.95 & 167 & 1.5 $\pm$ 0.02 & 6.28 & 116 & 1.12 $\pm$ 0.09 & 7.58  & 190 & 0.79 $\pm$ 0.04 & 7.13 \\
    \lbrack CRN2020\rbrack\ W1067 & KMOS & 339.59 & B0 Iab & 9.77 & 8.16 & 3 & 1.75 & 193 & 1.44 $\pm$ 0.03 & 6.14 & 71 & 1.18 $\pm$ 0.11 & 7.77 &  223 & 0.72 $\pm$ 0.04 & 4.79 \\
    Cl* Westerlund1 W15  & KMOS & 339.56 & O9 Ib & 11.22 & 9.64 & 3 & 1.73 & 111 & 1.48 $\pm$ 0.08 & 5.98 &  50 & 1.16 $\pm$ 0.18 & 8.84  & 234 & 0.77 $\pm$ 0.03 & 5.94 \\
    Cl* Westerlund1 W18 & KMOS & 339.55 & B0.5 Ia & 9.64 & 8.27 & 3 & 1.49 & 190 & 1.4 $\pm$ 0.04 & 6.90 & 52 & 1.12 $\pm$ 0.17 & 10.1 & 190 & 0.74 $\pm$ 0.04 & 7.22 \\
    Cl* Westerlund1 W24 & KMOS & 339.54 & O9 Iab & 10.3 & 8.6 & 3 & 1.86 & 154 & 1.5 $\pm$ 0.04 & 6.60 & 61  & 1.14 $\pm$ 0.15 & 9.51 & 163 & 0.75 $\pm$ 0.05 & 5.89 \\
    Cl* Westerlund1 W38 & KMOS & 339.55 & O9 Iab & 11.03 & 9.38 & 3 & 1.81 & 160 & 1.41 $\pm$ 0.03 & 5.71 & 59 & 1.13 $\pm$ 0.16 & 10.02 & 160 & 0.76 $\pm$ 0.06 & 10.08 \\
    Cl* Westerlund1 W43c & KMOS & 339.55 & O9 Ib & 10.92 & 9.25 & 3 & 1.83 & 180 & 1.41 $\pm$ 0.04 & 5.66 & 61 & 1.14 $\pm$ 0.17 & 10.04 & 165 & 0.76 $\pm$ 0.05 & 6.96 \\
    Cl* Westerlund1 W61b & KMOS &  339.53 & O9.5 Iab & 10.69 & 9.06 & 3 & 1.77 & 148 & 1.39 $\pm$ 0.06 & 5.76 &  80 & 1.15 $\pm$ 0.14 & 11.82 & 177 & 0.7 $\pm$ 0.05 & 6.89 \\
    LHO 1 & KMOS & 0.16 & O7-8Ia+/WNLh & 14.51 & 10.61 & 4 & 4.08 & - & - & - & 41 & 1.86 $\pm$ 0.27 & 9.51 & 123 & 1.28 $\pm$ 0.06 & 6.02 \\
    LHO 89 \footnotemark[4] & KMOS & 0.16 & O7.5-8.5If & 15.84 & 11.59 & 4 & 4.42 & - & - & - &  59 & 1.83 $\pm$ 0.12 & 8.77  & 143 & 1.28 $\pm$ 0.05 & 6.53 \\
    qF 134 (\textit{Pistol Star})\footnotemark[4] & KMOS & 0.15 & LBV & 11.79 & 7.74 & 4 & 3.87 & - & - & - & 98 & 1.78 $\pm$ 0.07 & 9.26  & 201 & 1.31 $\pm$ 0.03 & 6.50 \\
    qF 274 & KMOS & 0.17 & WN9 & 14.83 & 10.90 & 4 & 3.80 & - & - & - & 68 & 1.70 $\pm$ 0.14 & 8.84  & 279 & 1.28 $\pm$ 0.03 & 6.81 \\
    qF 406 & KMOS & 0.17 & O7-8Ia+ & 15.56 & 10.76 & 4 & 4.97 & - & - & - & 36 & 1.84 $\pm$ 0.25 & 9.14  & 147 & 1.43 $\pm$ 0.05 &  6.59 \\
    WR102e & KMOS & 0.15 & WC8-9 & 15.19 & 10.42 & 1 & 4.54 & - & - & - & 41 & 1.96 $\pm$ 0.25 & 9.74 & 72 & 1.45 $\pm$ 0.13 & 6.68 \\
    \lbrack DWC2011\rbrack\ 98 (\textit{G0.070}) \footnotemark[4] & GNIRS & 0.06 & O4-6If+ & 14.79 & 9.86 & 5 & 5.13 & 80 & 2.79 $\pm$ 0.13 & 6.00  &  50 & 1.62 $\pm$ 0.18 & 7.65  &  120 & 1.31 $\pm$ 0.15 & 5.66 \\
    \lbrack DWC2011\rbrack\ 23 (\textit{G0.071}) \footnotemark[4]  & GNIRS & 0.07 & O6-7 Ia+ & 14.98 & 10.78 & 5 & 4.37 & 40 & 3.23 $\pm$ 0.19 & 6.44  & 24 & 1.80 $\pm$ 0.24 & 10.92 & 126 & 1.4 $\pm$ 0.05 & 6.42 \\ 
    \lbrack DWC2011\rbrack\ 38 (\textit{G359925}) & GNIRS & 359.92 & O4Ia & 11.05 & 8.89 & 5 & 2.35 & 100 & 1.81 $\pm$ 0.1 & 5.90 & 60 & 1.01 $\pm$ 0.13 & 9.02 & 277 & 0.89 $\pm$ 0.02 & 5.31 \\
    \lbrack DWC2011\rbrack\ 36 & GNIRS & 359.97 & O4-6I & 15.13 & 11.36 & 5 & 3.96 & 50 & 2.42 $\pm$ 0.12 & 5.85  & 30 & 1.56 $\pm$ 0.3 & 9.61  & 45 & 1.24 $\pm$ 0.13 & 6.62 \\
    GCS 4 \footnotemark[5] & GNIRS & 0.16 & WC8/9d+OB & 14.53 & 7.23 & 4 & 4.89 & 54 & 3.43 $\pm$ 0.18 & 6.12  & 41 & 1.87 $\pm$ 0.21 & 8.90 & 200 & 1.52 $\pm$ 0.04 & 5.89 \\
    qF 256 & GNIRS & 0.16 & WN9 & 14.60  & 10.22 & 1 & 4.25 & 45 & 2.95 $\pm$ 0.15 & 6.12 &  38 & 1.62 $\pm$ 0.26 & 8.16 & 93 & 1.27 $\pm$ 0.07 & 5.69 \\
    qF 320 & GNIRS & 0.16 & WN9 & 15.18  & 10.82 & 4 & 4.62 & 52 & 3.01 $\pm$ 0.08 & 6.37  & 48 & 1.72 $\pm$ 0.15 & 9.30 & 121 & 1.32 $\pm$ 0.06 & 5.92 \\
    qF 362 \footnotemark[4] & GNIRS & 0.17 & LBV & 12.31 & 7.09 & 1 & 5.04 & 94 & 2.96 $\pm$ 0.06 & 6.23 & 46 & 1.92 $\pm$ 0.17 & 9.86 & 129 & 1.48 $\pm$ 0.03 & 6.26 \\
    TIC 126154878 \footnotemark[5]  & GNIRS & 0.53 & WN7-8h & 16.86 & 11.22 & 2 & 5.51 & 28 & 3.26 $\pm$ 0.31 & 7.52  & 32 & 2.29 $\pm$ 0.27 & 10.62 & 53 & 1.6 $\pm$ 0.15 & 6.13 \\
    WR102ka & GNIRS & 0 & WN10h & 12.98 & 8.83 & 1 & 4.20 & 93 & 3.44 $\pm$ 0.07 & 6.43  & 93 & 1.72 $\pm$ 0.11 & 7.57  & 140 & 1.28 $\pm$ 0.06 & 4.13 \\
    TIC 195381536 (\textit{J2039}) & GIANO-B & 81.82 & B0I & 7.34 &  5.82 & 1 & 1.65 & 132 & 1.43 $\pm$ 0.01 & 4.70 & 183 & 0.64 $\pm$ 0.01 & 6.15  & 223 & 0.6 $\pm$ 0.01 & 3.78 \\
    TIC 109609771  \footnotemark[4] & GIANO-B & 12.79 & LBV & 8.92 & 6.96 & 1 & 1.78 & 85 & 2.01 $\pm$ 0.03 & 5.47 & 56 & 1.02 $\pm$ 0.05 & 7.02 & 178 & 0.45 $\pm$ 0.02 & 3.87 \\ 
    TIC 175996914 (\textit{GAL026}) \footnotemark[4] & GIANO-B & 26.46 & LBV & 7.99 & 5.61 & 1 & 2.20 & 163 & 2.13 $\pm$ 0.02 & 6.54  & 143 & 1.1 $\pm$ 0.07 & 7.35  & 152 & 0.95 $\pm$ 0.01 & 4.42 \\
    CygOB2 7 & GIANO-B & 80.24 & O3If* & 7.25 & 6.61 & 1 & 0.85 & 143 & 0.92 $\pm$ 0.01 & 4.37  & 97 & 0.48 $\pm$ 0.04 & 4.9 & 170 & 0.57 $\pm$ 0.01 & 4.42 \\
    CygOB2 11 & GIANO-B & 80.56 & O5.5Ifc & 6.65 & 5.99 & 1 & 0.86 & 162 & 1.18 $\pm$ 0.01 & 4.56 & 100 & 0.58 $\pm$ 0.05 & 4.51 & 225 & 0.57 $\pm$ 0.01 & 3.95 \\
    CygOB2 12 \footnotemark[4] & GIANO-B & 80.10 & B3-4Ia+ & 4.66 & 2.7 & 1 & 2.05 & 208 & 1.21 $\pm$ 0.01 & 4.62 & 113 & 0.55 $\pm$ 0.03 & 5.7  & 240 & 0.58 $\pm$ 0.01 & 4.06 \\
    WR 105 (\textit{NS4})  & GIANO-B & 6.52 & WN9h & 7.03 & 5.73 & 1 & 1.17 & 120 & 1.06 $\pm$ 0.01 & 5.41  & 64 & 0.57 $\pm$ 0.03 & 6.67 & 145 & 0.42 $\pm$ 0.01 & 3.75 \\
    TIC 196042957 (\textit{MGE042})\footnotemark[4] & GIANO-B & 42.07 & LBV & 9.66 & 6.85 & 1 & 2.63 & 108 & 1.76 $\pm$ 0.01 & 5.63  & 107 & 0.84 $\pm$ 0.03 & 6.77 & 112 & 0.81 $\pm$ 0.02 & 4.13 \\
    HD190603 \footnotemark[4]  & GIANO-B & 69.40 & B1.5Ia+ & 4.42 & 4.04 & 1 & 0.48 & 121 & 0.49 $\pm$ 0.01 & 4.46 & 66 & 0.35 $\pm$ 0.06 & 5.34 & 101 & 0.18 $\pm$ 0.02 & 3.33 \\
    IRAS 20298+4011 (\textit{G79})\footnotemark[4] & GIANO-B & 79.28 & B:I[e] & 6.91 & 4.33 & 1 & 2.68 & 173 & 0.76 $\pm$ 0.01 & 4.34 & 71 & 0.50 $\pm$ 0.04 & 6.35 & 270 & 0.48 $\pm$ 0.01 & 4.12 \\
    IRAS 19425+2411 (\textit{MN112})\footnotemark[4] & GIANO-B & 60.45 & LBV & 8.86 & 7.42 & 1 & 1.26 & 102 & 0.99 $\pm$ 0.02 & 4.83 & 105 & 0.59 $\pm$ 0.03 & 7.25  & 182 & 0.40 $\pm$ 0.02 & 3.81 \\
    \lbrack DWC2011\rbrack\ 103 & Xshooter & 0.05 & OI & 13.33 & 9.02 & 1 & 4.34 & 71 & 2.3 $\pm$ 0.06 & 5.42 & 30 & 1.47 $\pm$ 0.19 & 7.92 & 110 & 1.17 $\pm$ 0.05 & 5.09 \\
    \lbrack DWC2011\rbrack\ 145 & Xshooter & 359.73 & O7-O8If & 9.90 & 7.97 & 1 & 2.10 & 104 & 1.59 $\pm$ 0.04 & 5.15 & 58 & 0.89 $\pm$ 0.13 & 7.64 & 81 & 0.81 $\pm$ 0.04 & 4.22 \\
    TIC 116433010 (\textit{B164})& Xshooter & 15.07 & O6V & 10.01 & 8.76 & 1 & 1.41 & 96 & 0.99 $\pm$ 0.03 & 5.76 & 52 & 0.60 $\pm$ 0.09 & 8.28 & 72 & 0.49 $\pm$ 0.05 & 5.02 \\
    TIC 116432629 & Xshooter & 15.07 & O9.7V & 10.45 & 9.18 & 1 & 1.44 & 115 & 0.96 $\pm$ 0.03 & 6.33  & 79 & 0.56 $\pm$ 0.06 & 8.02 & 182 & 0.42 $\pm$ 0.02 & 4.76 \\ 
    HD80077 & Xshooter & 271.62 & B2Ia+e & 4.44  & 3.63 & 1 & 0.92 & 84 & 0.64 $\pm$ 0.06 & 4.00 & 59 & 0.43 $\pm$ 0.08 & 6.53 & 130 & 0.37 $\pm$ 0.06 & 4.32 \\
    Cl* Westerlund1 W7  \footnotemark[4] & Xshooter & 339.56 & B5 Ia+ & 6.94 & 5.41 & 3 & 1.59 & 100 & 1.53 $\pm$ 0.12 & 5.34 & 46 & 1.22 $\pm$ 0.15 & 8.57 & 150 & 0.87 $\pm$ 0.06 & 5.34 \\
    Cl* Westerlund1 W33 \footnotemark[4] & Xshooter & 339.55 & B5 Ia+ & 6.88 & 5.29 & 3 & 1.64 & 100 & 1.48 $\pm$ 0.12 & 5.56 & 48 & 1.18 $\pm$ 0.14 & 8.57 & 115 & 0.85 $\pm$ 0.08 & 5.57 \\
    Cl* Westerlund1 W57a \footnotemark[4] & Xshooter & 339.53 & B4 Ia & 8.13 & 6.93 & 3 & 1.28 & 105 & 1.56 $\pm$ 0.12 & 5.34 & 66 & 1.10 $\pm$ 0.11 & 8.05 & 200 & 0.8 $\pm$ 0.06 & 5.22 \\
    HD10747 & Xshooter & 298.86 &  B2V & 8.44 & 8.59 & 1 & 0.05 & 117 & 0.07 $\pm$ 0.05 & - & 53 & 0.05 $\pm$ 0.2  & -  & 99 & 0.08 $\pm$ 0.01 & - \\
    HD48434 & Xshooter & 208.54 & B0II & 5.94 & 5.98 & 1 & 0.12 & 98 & 0.06 $\pm$ 0.03 & - & 42 & 0.02 $\pm$ 0.1  & -  & 81 & 0.05 $\pm$ 0.03 & - \\
    HD154445 & Xshooter & 19.29 & B1V & 5.31 & 5.29 & 1 & 0.17 & 126 & 0.09 $\pm$ 0.05 & - & 62 & 0.01 $\pm$ 0.04 & -  & 72 & 0.02 $\pm$ 0.03 & - \\
    HD203532 & Xshooter & 309.45 & B3.5V & 6.06 & 6.02 & 1 & 0.14 & 137 & 0.03 $\pm$ 0.02 & - & 56 & 0.03 $\pm$ 0.04 & -  & 198 & 0.01 $\pm$ 0.01 & - \\
    HD209008 & Xshooter & 65.80 & B3III & 6.19 &  6.23 & 1 & 0.12 & 111 & 0.05 $\pm$ 0.1 & -  & 49 & 0.05 $\pm$ 0.08 & -  & 114 & 0.05 $\pm$ 0.02 & - \\

\end{longtable}

\begin{minipage}{0.95\linewidth}
\vspace{12mm}
\renewcommand{\thempfootnote}{\arabic{mpfootnote}}
\footnotetext[1]{Angle relative to that toward Sgr A* in the GC.}
\footnotetext[2]{Reference for photometry: (1) - \citet{Cutri_2mass_2003}, (2) - \citet{VISTA_2012}, (3) - \citet{Gennaro_2017}, (4) - \citet{Clark_2018}, (5) - \citet{Mauerhan_2010}.}
\footnotetext[3]{The intrinsic colours of the stars are derived from specific sources: OB types are derived from \citet{Wegner_1994}, WR stars are based on \citet{Crowther_2006}. LBVs are approximated to the intrinsic colour of WN11h stars, reflecting their spectral and luminosity characteristics.}
\footnotetext[4]{Sources whose $\lambda$15272\,Å DIB is contaminated and measurements were obtained using the Gaussian fit detailed in Appendix \ref{sec:Appendix_A}.}
\footnotetext[5]{Sources where the dust contribution to the $E(J-K)$ colour index has been removed following \citet{Najarro_GC_2017}.}
\end{minipage}

\end{landscape}
\clearpage
\twocolumn

\subsection{DIB profiles}

\citet{ramirez2018diffuse}, \citet{dahlstrom2013anomalous}, and \citet{oka2013anomalous} documented various optical DIB profile types.
\citet{rawlings2014high} found that the profiles of two near-infrared DIBs, at $\lambda$11797\,Å and $\lambda$13177\,Å are asymmetric with elongated red wings. 
\citet{galazutdinov2017infrared} found a similar asymmetry of the $\lambda$14680\,Å DIB that they observed on two sightlines.
As shown in Fig. \ref{fig:Gaussian_fits} the shapes of $\lambda$13177\,Å and $\lambda$14680\,Å DIBs are asymmetric, with extended red wings consistent with the findings of \citet{rawlings2014high} and \citet{galazutdinov2017infrared}. 
On the other hand, the $\lambda$15272\,Å DIB consistently has the shape of single Gaussian, even for the high resolution observations from TNG-GIANO-B (see the J2039 and NS4 spectra in Fig. \ref{fig:Gaussian_fits}).
The symmetry of this DIB has previously been reported \citep{galazutdinov2017infrared}.
This diversity suggests either distinct carrier structures or varying environmental conditions at the origin of these DIBs.

\subsection{Rest wavelength of detected features}

Adopting the methodology used by \citet{elyajouri2017near}, we utilize the established rest wavelength value in vacuum of the $\lambda$15272\,Å DIB determined by \citep{zasowski2014mapping}, as a reference for accurately determining the rest wavelengths of the other four DIBs ($\lambda$13177\,Å, $\lambda$14680\,Å, $\lambda$14795\,Å, and $\lambda$14850\,Å) studied in this paper.

Our approach involves two steps: 
We initially apply a flux-weighted method to determine the reference wavelength of each DIB
in individual spectra. 
This method calculates the reference wavelength by weighting the intensity difference between the continuum (normalized to 1) and the flux at each point of the DIB profile. Next, we utilize the technique of \citet{elyajouri2017near}, who calculates the Doppler shifts of the selected DIBs. 
This approach assumes that the Doppler shifts of all DIBs are the same for a given line of sight. 
We determine the difference between the observed wavelength of the $\lambda$15272\,Å DIB and the established value of \citet{zasowski2014mapping} for each spectrum. 
This approach ensures the uniformity of the calculated shifts, allowing for a precise determination of the central wavelengths. 
Using this approach, we achieve robust determinations of the rest wavelengths of the remaining DIBs.
The central wavelengths of the studied DIBs and their uncertainties are given in Table \ref{tab:DIBs_rest_wave} determined from the standard deviation of the values obtained from different spectra.

\begin{table}
\centering
\caption{Vacuum wavelengths of four NIR DIBs. The rest wavelengths values are determined using as a reference the rest wavelength value of \citet{zasowski2014mapping} for $\lambda$15272\,Å DIB.}{
\begin{tabular}{ll}
\hline
\multicolumn{1}{c}{DIB}      & \multicolumn{1}{c}{Rest wavelength (Å)} 
\\ 
\hline
\hline
\multicolumn{1}{c}{13177} & \multicolumn{1}{c}{13181.07$\pm$0.16}                  \\
\multicolumn{1}{c}{14680} & \multicolumn{1}{c}{14681.28$\pm$0.18}  \\
\multicolumn{1}{c}{14795} & \multicolumn{1}{c}{14798.12$\pm$0.22}                  \\
\multicolumn{1}{c}{14850} & \multicolumn{1}{c}{14852.62$\pm$0.31}                  \\

\hline
\end{tabular}%
}
\label{tab:DIBs_rest_wave}
\end{table}

\section{Discussion}
\label{sec:Discussion}


Previous studies of optical DIBs have revealed strong correlations between the EWs of many of them and fundamental interstellar parameters, such as reddening and line of sight gas column density among others \citep{Herbig_1993, Friedman_2011, Kos_2013}. 
Below we examine the behaviour of the EWs of the  $\lambda$13177\,Å, $\lambda$14680\,Å and $\lambda$15272\,Å DIBs as  functions of reddening for our large sample of sightlines.
We also investigate the ratios of the EWs of these DIBs to test for any trends as a function of reddening and sightline.
Our sample includes three different clusters and an OB association whose members suffer a moderate range of extinctions.
This allow us to investigate whether the DIBs carriers are present in the local surroundings of the clusters.  


\subsection{Environmental Impact on DIBs}
\label{sec:DIBs_reddening}

\begin{figure*}
	\includegraphics[width=\textwidth]{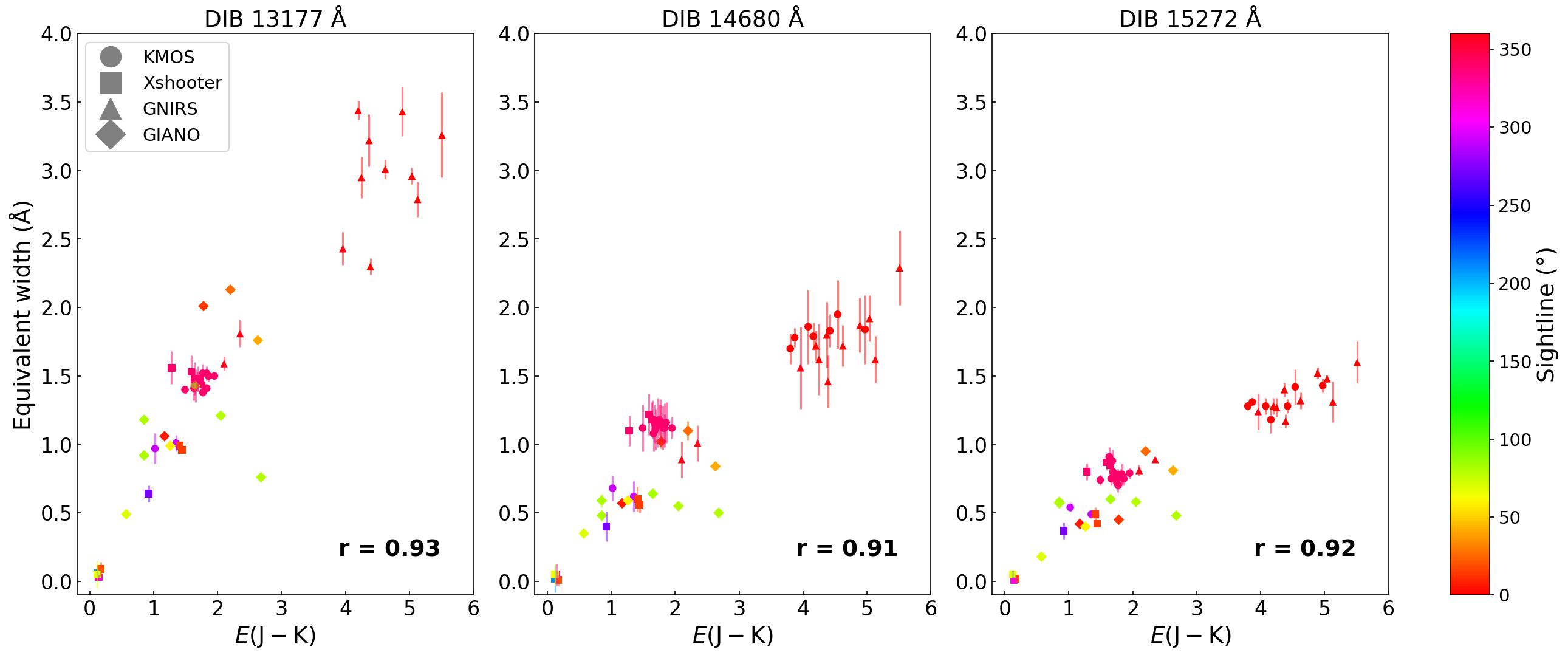}
    \caption{Equivalent width in\,Å vs. $E(J-K)$ index for all observed sightlines for the $\lambda$13177\,Å, $\lambda$14680\,Å, and $\lambda$15272\,Å DIBs.}
    \label{fig:DIBs_ew_reddening}
\end{figure*}

In our analysis of NIR DIBs, we make use of the correlation coefficient classification scheme originally proposed by \citet{Moutou_1999} for optical DIBs. 
According to this scheme, a Pearson correlation coefficient (r) less than 0.7 is considered weak, between 0.7 and 0.95 as good, and above 0.95 as strong. 
This classification aligns with that proposed by \citet{Fan_2017} and \citet{Friedman_2011}. 
Both papers point out that an r-value in the range of 0.86 to 0.88 or higher may indicate a physical relationship between the studied quantities. 
For DIBs arising from the same carriers or those whose carriers formed in the presence of a common third carrier, the correlation coefficients are expected to approach unity.
A fairly good correlation between DIBs is expected, simply due to the progressive accumulation of absorbing material with increasing line of sight distance. 
Consequently, high values of Pearson's correlation coefficient (e.g. r > 0.8) are crucial for identifying actual physical relationships between DIBs.

Figure \ref{fig:DIBs_ew_reddening} displays the EWs of the $\lambda$13177\,Å, $\lambda$14680\,Å, and $\lambda$15272\,Å DIBs as functions of the reddening index $E(J-K)$ for all lines of sight studied in this paper. 
After reviewing the published NIR photometry, we have adopted a conservative reddening error estimate of 0.06 mag.
We found that detecting these DIBs is challenging for stars with $E(J-K)$ less than 1 mag.
On all observed lines of sight, the EW of the $\lambda$13177\,Å DIB is consistently larger than that of the $\lambda$15272\,Å DIB.
\citet{cox2014vlt} found a similar result in their sample.
This suggests that the carrier for the former is either more abundant or has a higher absorbance per molecules, or both.
The EWs of the $\lambda$14680\,Å and $\lambda$15272\,Å DIBs are more nearly equal on these lines of sight, but on most of them the EW of the $\lambda$14680\,Å DIB is slightly greater.
The EWs of each of these three DIBs are strongly correlated with $E(J-K)$ in our diverse sample.
Although the scatter in the measurements increases for $E(J-K)$ > 3 mag, there is a small decrease in slope discernible above $E(J-K)$ = 3 mag.
However, as discussed below, slightly different behaviours of these DIBs on different groups of sightlines are at least partly responsible for the apparent changes in slope.

For each of these DIBs, the Pearson correlation coefficient between the EW and reddening is $\approx$ 0.9. This is in line with the findings reported by \citet{cox2014vlt}  and \citet{smoker2022high} for the $\lambda$13177\,Å and $\lambda$15272\,Å DIBs. 
However, this correlation coefficient is significantly higher than the reported by \citet{Hamano_2015}.
These results support the widely accepted hypothesis that like most other DIBs, the strength of each of these DIBs is closely tied to the column density of interstellar material on its sightline, and extends that relationship to much larger reddening than previous studies.
We also clearly establish the interstellar origin of the $\lambda$14680\,Å DIB, which had only been observed previously on two sightlines.

\begin{figure}
	\includegraphics[width=\columnwidth]{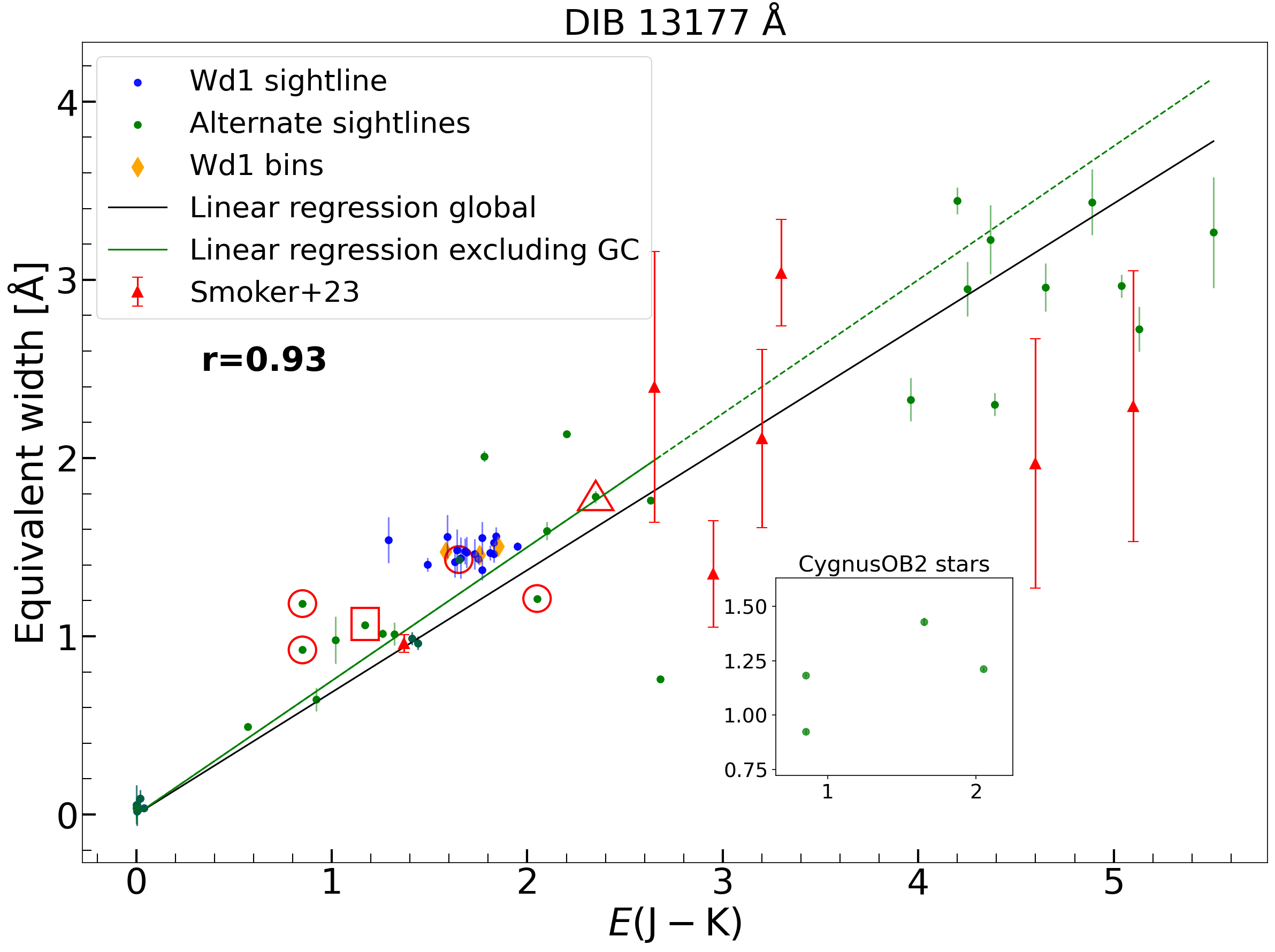}
    \caption{Correlation between the equivalent width of the $\lambda$13177\,Å DIB and the reddening for all objects in the study. The observed sources are represented as green dots while Wd1 cluster members are represented as blue dots. The data points of the CygOB2 association are marked as red circles, while NS4 star is marked with a red square and G359925 star is highlighted using a red triangle. A linear regression including all sources is represented in black. A linear regression excluding GC sources is represented in green. The bins aggregating the Wd1 sources are represented as orange diamonds. The correlation coefficient is noted on the plot. Highly reddened sources previously studied by \citet{smoker2022high} are included as red triangles}.
    \label{fig:1.31_EW_reddening_corr}
\end{figure}

\begin{figure}
	\includegraphics[width=\columnwidth]{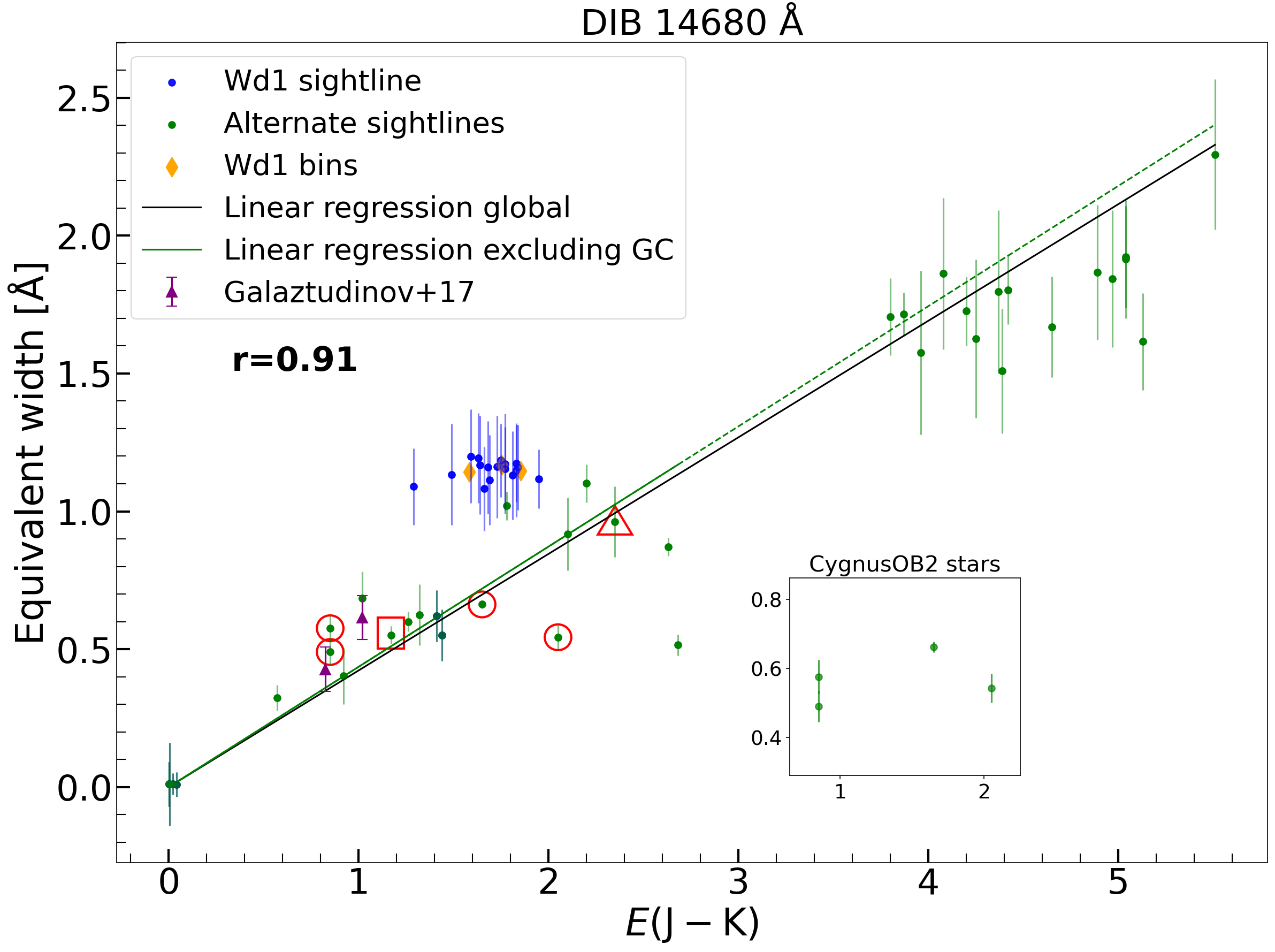}
    \caption{Analogue to Fig. \ref{fig:1.31_EW_reddening_corr}, but for the $\lambda$14680\,Å DIB. See Fig. \ref{fig:1.31_EW_reddening_corr} caption for symbols and labels.
    Two sources studied by \citet{galazutdinov2017infrared} are included as purple triangles.}
    \label{fig:1.46_EW_reddening_corr}
\end{figure}

\begin{figure}
	\includegraphics[width=\columnwidth]{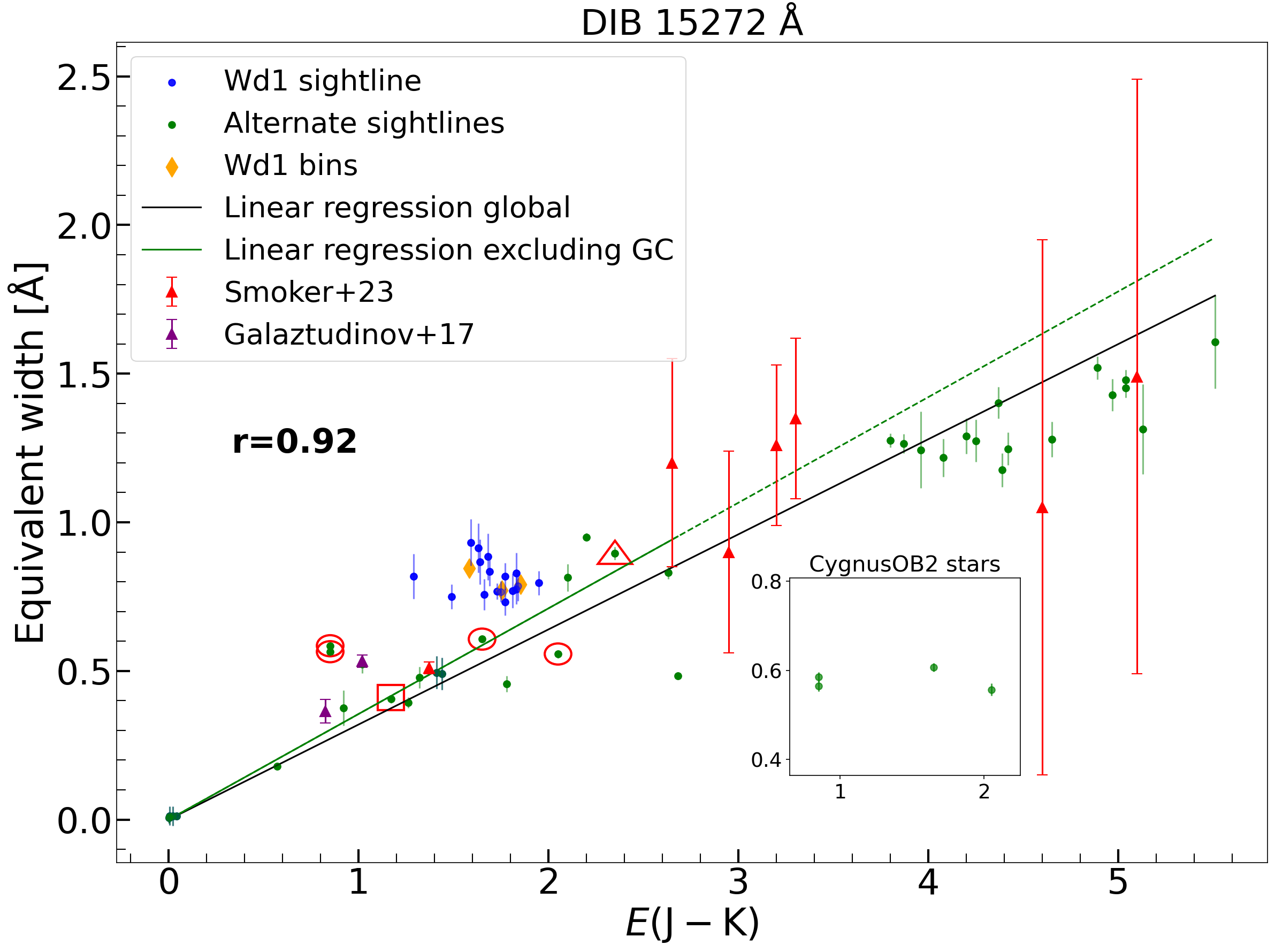}
    \caption{Analogue to Fig. \ref{fig:1.31_EW_reddening_corr}, but for the $\lambda$15272\,Å DIB. See Fig. \ref{fig:1.31_EW_reddening_corr} caption for symbols and labels. Two sources studied by \citet{galazutdinov2017infrared} are included as purple triangles.}

    \label{fig:1.52_EW_reddening_corr}
\end{figure}

More detailed examination of the data reveals differences in the EW vs $E(J-K)$ relation on different sightlines. 
This is illustrated in Figs. \ref{fig:1.31_EW_reddening_corr}, \ref{fig:1.46_EW_reddening_corr} and \ref{fig:1.52_EW_reddening_corr}. 
To check consistency with previous studies, we have included EWs values of heavily reddened stars previously studied by \citet{smoker2022high} and two sources studied by \citet{galazutdinov2017infrared}.
To calculate the $E(J-K)$ values of these stars we followed the methodologies described by \citet{Wegner_1994} for OB stars and by \citet{Minniti_2020} for Cepheids.
In the case of the \citet{smoker2022high} dataset we reduced and analysed the spectra from the archive and made our own measurements, which were in line with the reported values.
Additionally, we applied the correction for hydrogen contamination present in the $\lambda$15272\,Å DIB following the methodology detailed in Appendix \ref{sec:Appendix_A} to obtain the values included in Fig. \ref{fig:1.52_EW_reddening_corr}.
Despite some of these sources having a low S/N, the values align with the trends observed in the sightlines we have examined.
For all three DIBs the measured EWs in the Wd1 cluster lie above the regression.
In particular, the $\lambda$14680\,Å DIB in Wd1 displays a larger deviation than the other two DIBs.
The EWs observed towards the highly reddened GC stars are lower than expected based on the trends observed at lower reddening values. 
This could be indicative of the more extreme conditions inhibiting the formation of DIBs carriers in the regions closer to the GC.
To investigate this scenario more quantitatively, we derived two linear regressions for each DIB, one including and one excluding those objects with $E(J-K)$ > 3.
To prevent the high number of sources from the Wd1 cluster from overly biasing the new regression calculation, we grouped them into three bins.
The resulting linear regressions (see Figs. \ref{fig:1.31_EW_reddening_corr}, \ref{fig:1.46_EW_reddening_corr} and \ref{fig:1.52_EW_reddening_corr}) confirm  the higher EWs in $\lambda$14680\,Å and $\lambda$15277\,Å DIBs for Wd1 line of sight as well as the lower EWs on GC sightlines, consistent with carrier formation being inhibited in the GC.
This pattern suggests that molecular clouds along the Wd1 line of sight have certain physical properties that somewhat favour the prominence of the $\lambda$14680\,Å DIB carrier, and appear to favour to a lesser extent the carriers of other DIBs. 
However, explaining this difference is not straightforward.
Since Wd1 is located at a smaller galactocentric radius than most of the sources studied, a higher metallicity could contribute to a larger equivalent width of the DIBs.
This is challenged by observations such as those of NS4 and G359925 (highlighted with a red square and a red triangle respectively in Figs. \ref{fig:1.31_EW_reddening_corr}, \ref{fig:1.46_EW_reddening_corr} and \ref{fig:1.52_EW_reddening_corr}), which are closer to the GC than Wd1, but does not show this increased DIB width. 
This indicates that factors other than mere proximity to the GC may be involved.
Such a scenario suggests that one of the possible options is the existence of a radial-azimuthal behaviour in metallicity, rather than a strictly radial pattern, as suggested by \citet{Davies_2009}.
Additionally, another factor to consider in the observed deviations of the EW of the DIBs could be related to the density of the molecular clouds encountered along different sightlines. 
\citet{Elyajouri_L_2017} reported how denser clouds can exhibit an effect where the EWs of the DIBs decrease relative to the reddening more than expected. 
Although this effect is not pronounced across all the sightlines we studied, the relatively higher EWs observed in the Wd1 line of sight could be indicative of less dense cloud present in this sightline compared to others.


Figs. \ref{fig:1.31_EW_reddening_corr},  \ref{fig:1.46_EW_reddening_corr} and  \ref{fig:1.52_EW_reddening_corr} illustrate that the star NS4, located along the line of sight to the GC; (see Fig.\ref{fig:mapa_galaxia}) exhibits EWs in all DIBs that align with the $E(J-K)$ value, adhering to the linear regression.
However, considering its distance of 4 kpc, as determined by \citet{Bailer_Jones_21}, it implies that the interstellar material along this sightline towards the GC is not uniformly distributed. 
The sightlines toward the GC traverse three spiral arms and a part of the Central Molecular Zone (CMZ), a region of radius $\sim$150 pc centred on Sgr A*, which contains large quantities of both dense and diffuse molecular gas (\citealt{Morris_1996}; \citealt{Oka_2019}).
In particular, the spiral arms are anticipated to contain a significant concentrations of interstellar material, including dust and DIB carriers. 
Therefore, based on the EW and reddening observed towards NS4, it can be inferred that the concentration of interstellar material from NS4 to the GC may be at least twice as high as in the initial 4 kpc leading up to NS4.
However, since the intervening spiral arms are patchy, NS4's sightline could be anomalous, crossing through relative minima in spiral arm gas and dust.
Thus, further investigation of $E(J-K)$ values for stars at similar distances and lines of sight as NS4 is required.

A noteworthy result, visible in Figs. \ref{fig:1.31_EW_reddening_corr},  \ref{fig:1.46_EW_reddening_corr} and  \ref{fig:1.52_EW_reddening_corr}, is the nearly constant EWs of both the $\lambda$14680\,Å and the $\lambda$15272\,Å DIBs on sightlines to stars in the CygOB2 association (marked with red circles and highlighted using a smaller graph in the lower right area of the figures), despite large differences in reddening toward them.
A similar behaviour has been previously reported in CygOB2 in the case of the $\lambda$10504\,Å DIB by \citet{Hamano_2016}.
This suggests that the carriers responsible for both DIBs are not present within the cluster and thus are not directly related to the dust within the association that causes the extra reddening to different association members, but rather to foreground clouds in front of the entire CygOB2 association.
In contrast, the EW of the $\lambda$13177\,Å DIB displays significant differences among the CygOB2 stars that correlate with differences in reddening, indicating that its carrier is present within the association and pointing to a different carrier for this DIB than for the other two, both of which may be more vulnerable to the environmental conditions within the CygOB2 association.


Our findings reveal a consistency in the FWHM of the $\lambda$15272\,Å DIB for stars within the GC with the findings of \citet{geballe2011infrared}.
They observed a similar Doppler velocity broadening in the $\lambda$15272\,Å DIB towards GCS3-2, likely attributable to the different radial velocities of the three spiral arms along this sightline \citep{Oka_2019}.
Furthermore, our measured FWHMs for stars in less reddened regions are in agreement with values in the literature.
Specifically, stars in CygOB2 show similar FWHMs as those reported by \citet{cox2014vlt} for the $\lambda$13177\,Å and the $\lambda$15272\,Å DIBs, observed in lines of sight less affected by Doppler broadening.
This consistency between the different regions reinforces the interpretation that the broadening of the DIBs towards the GC is predominantly due to Doppler effects.




\subsection{Ratios between EWs of the three DIBs}
\label{sec:Ratios}

\begin{figure*}
	\includegraphics[width=\textwidth]{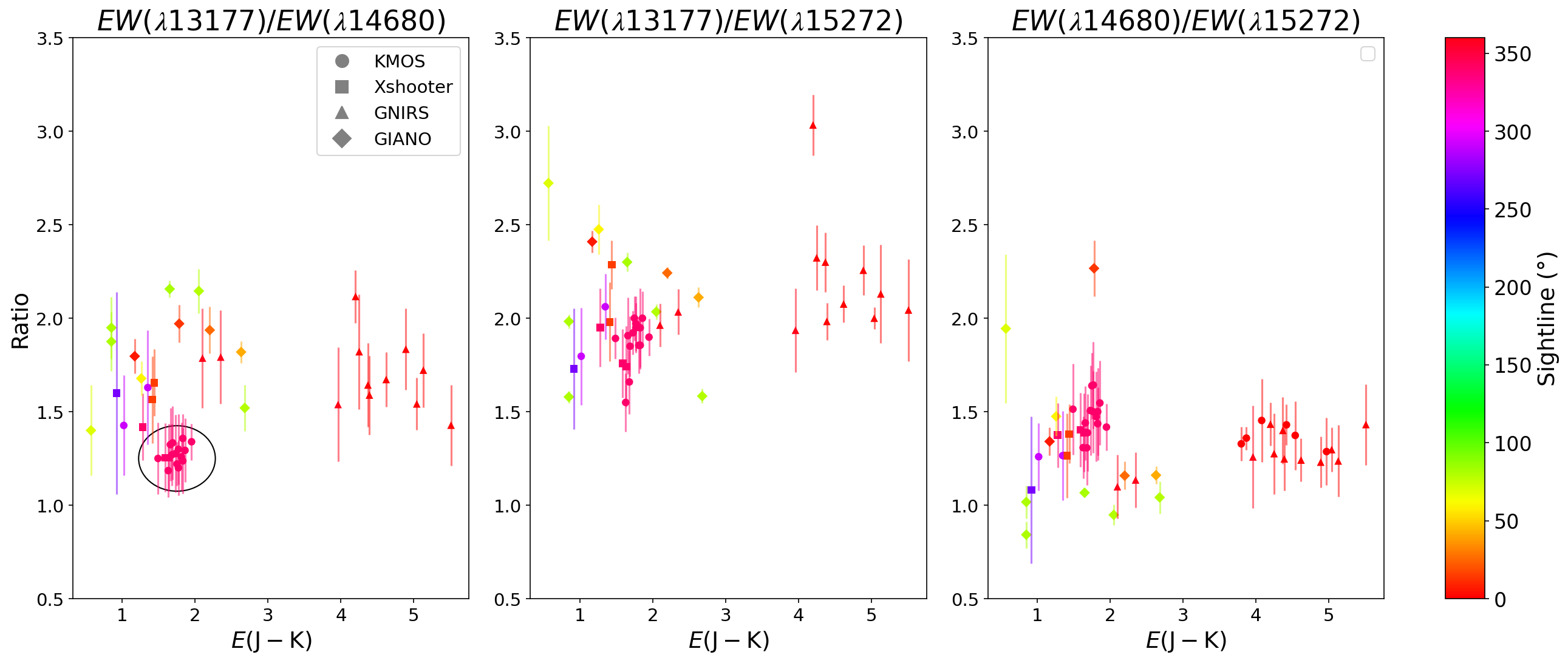}
    \caption{Ratios of EWs of pairs of DIBs vs. $E(J-K)$ index for the studied objects, presented in three subplots. DIBs pairs are indicated at the top of each panel. In the left panel the Wd1 stars are enclosed in a black circle. Stars in CygOB2 are denoted by green dots.}
    \label{fig:Ratios_DIBs_JK}
\end{figure*}

Figure \ref{fig:Ratios_DIBs_JK} contains plot of the ratios of pairs of DIB EWs vs. reddening. 
The nearly constant value of each ratio ($\lambda$13177/$\lambda$14680, $\lambda$13177/$\lambda$15272 and $\lambda$14680/$\lambda$15272) across varied Galactic environments suggests that the carriers of these DIBs are similarly distributed along the examined sightlines.
The largest deviations from this are for stars in the CygOB2 association and Wd1 cluster as discussed in Sect. \ref{sec:DIBs_reddening}. 

\subsection{DIBs correlations and implications}
\label{sec:DIBs_correlation}

\begin{figure}
	\includegraphics[width=\columnwidth]{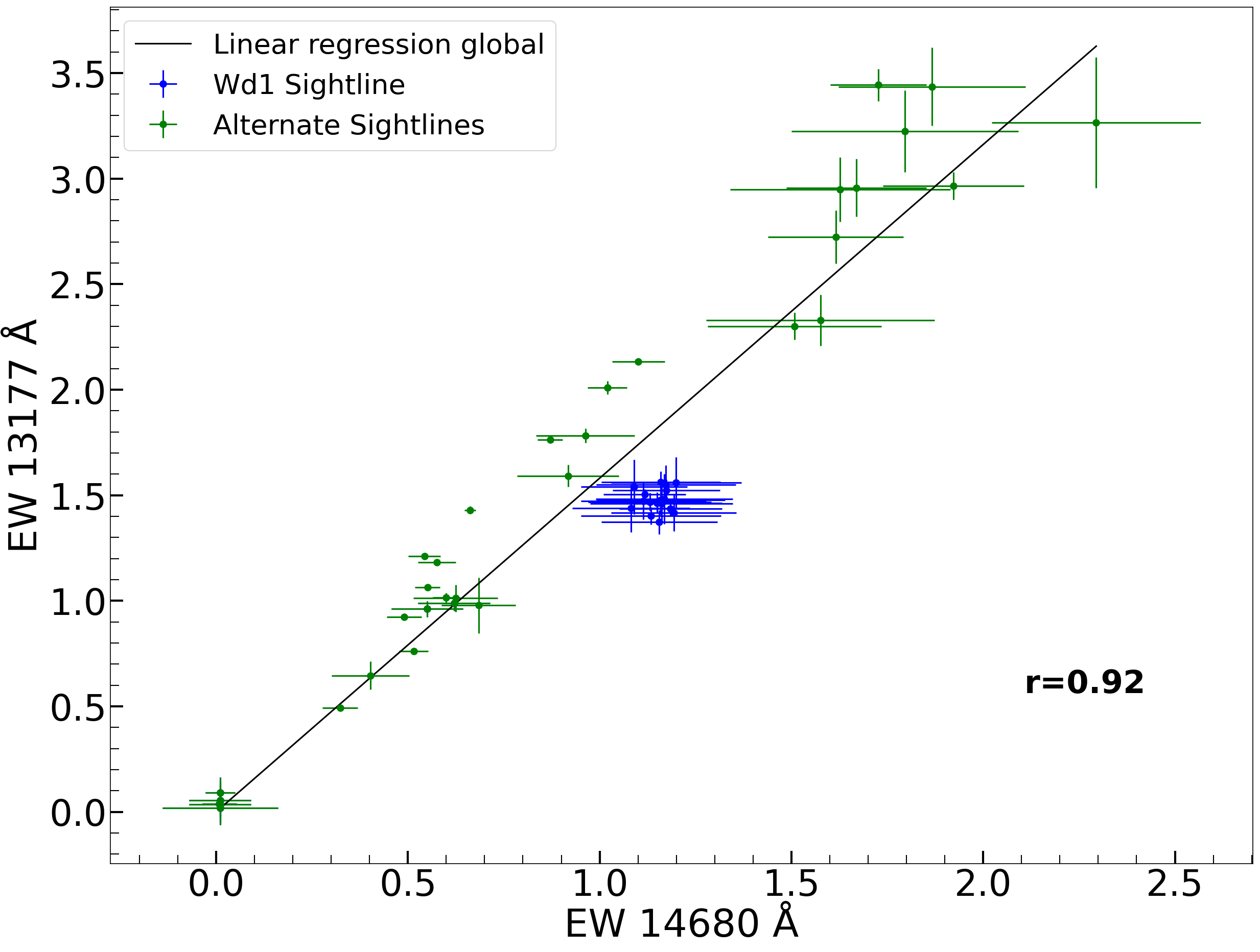}
    \caption{Correlation between the EWs of the $\lambda$13177\,Å and $\lambda$14680\,Å DIBs for all objects in our study. The observed sources are represented as green dots while Wd1 cluster members are represented as blue dots. A linear regression including all sources is represented in black. The correlation coefficient is noted on the plot.}
    \label{fig:Correlacion_EW_1.46775_1.31775}
\end{figure}

\begin{figure}
	\includegraphics[width=\columnwidth]{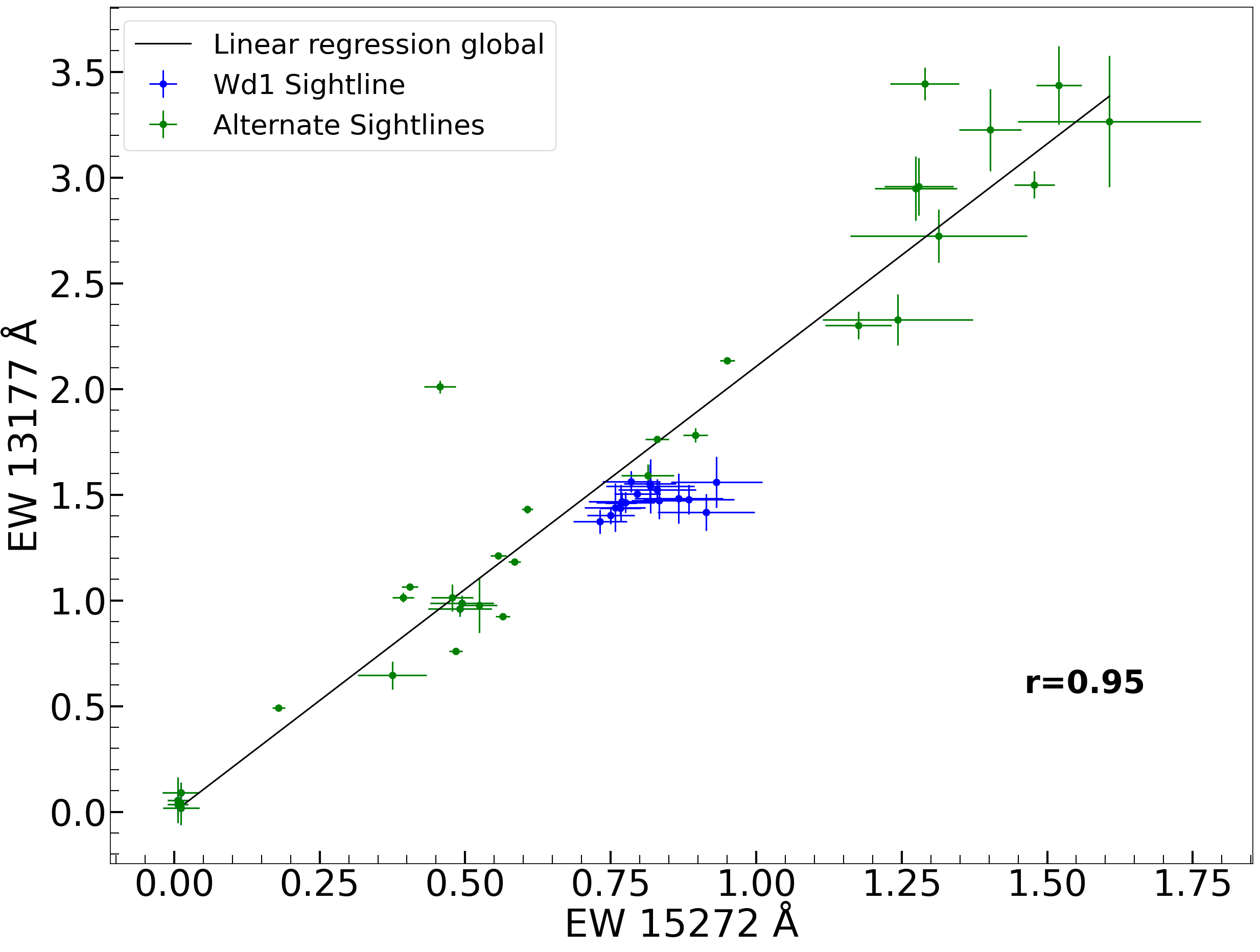}
    \caption{Analogue to Fig. \ref{fig:Correlacion_EW_1.46775_1.31775}, but for the correlation between $\lambda$13177\,Å and $\lambda$15272\,Å DIBs. See Fig. \ref{fig:Correlacion_EW_1.46775_1.31775} caption for symbols and labels.}
    \label{fig:Correlacion_EW_1.52695_1.31775}
\end{figure}

\begin{figure}
	\includegraphics[width=\columnwidth]{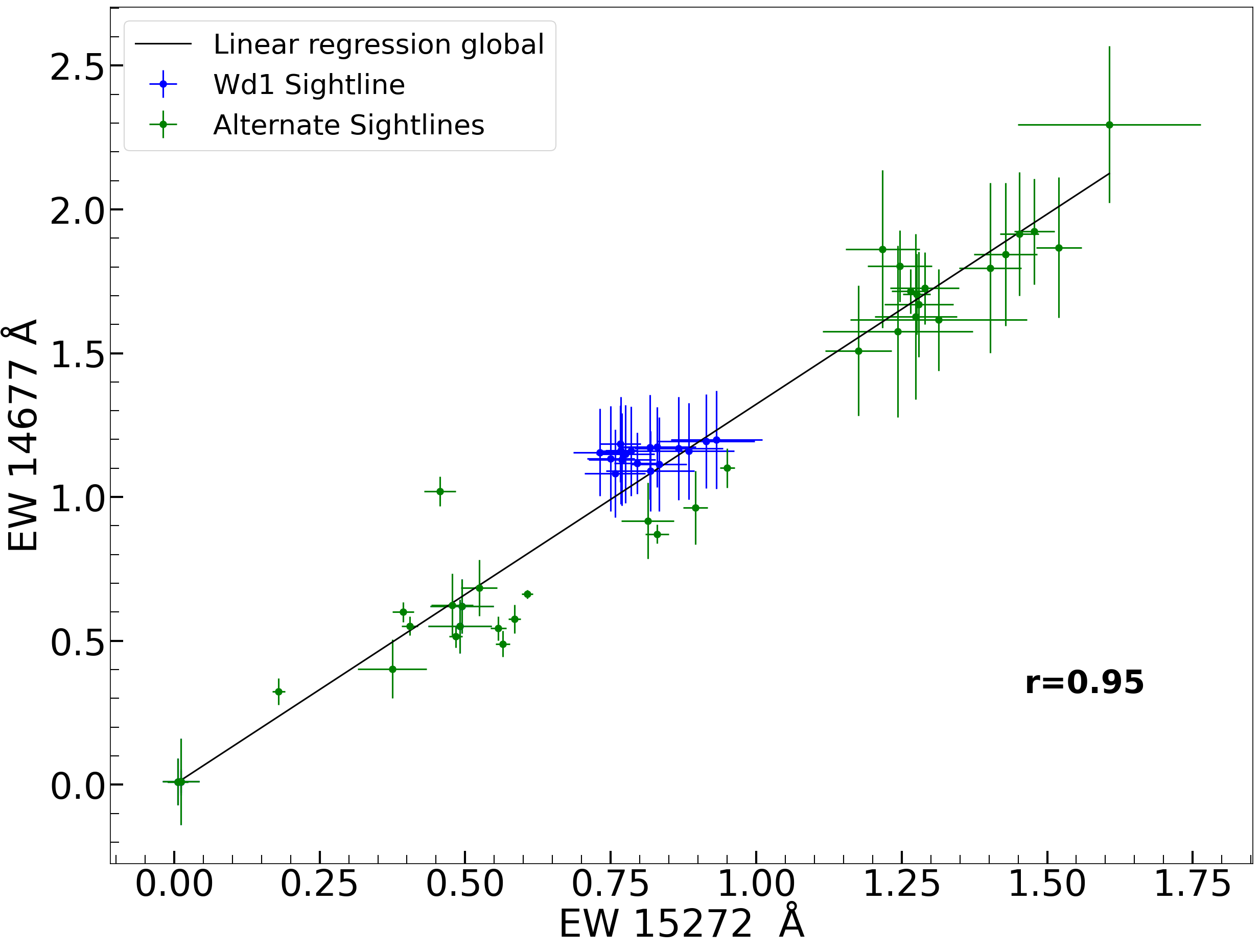}
    \caption{Analogue to Fig. \ref{fig:Correlacion_EW_1.46775_1.31775}, but for the correlation between $\lambda$14680\,Å and $\lambda$15272\,Å DIBs. See Fig. \ref{fig:Correlacion_EW_1.46775_1.31775} caption for symbols and labels.}
    \label{fig:Correlacion_EW_1.52695_1.46775}
\end{figure}

Correlation plots of the EWs of pairs of DIBs are presented for each pair of DIBs in Figs. \ref{fig:Correlacion_EW_1.46775_1.31775}, \ref{fig:Correlacion_EW_1.52695_1.31775}, and \ref{fig:Correlacion_EW_1.52695_1.46775}.  
Pearson coefficients for these pairs range from 0.92 to 0.95 (see Table \ref{tab:DIB Correlation Coefficients}), indicating robust correlations across the different sightlines. 
Yet, none of these correlation coefficients are sufficiently close to unity to suggest shared carriers or formation influenced by a common carrier.

It is evident in Fig. \ref{fig:Correlacion_EW_1.46775_1.31775} that the line of sight to Wd1 fall significantly below the linear regressions, unlike the remainder of the sightlines. 
This suggests that on at least part of the sightline toward the Wd1 cluster the chemistry in the diffuse interstellar medium differs from the average behaviour elsewhere. 
Nevertheless, the near constancy of the ratios of all three pairs of EWs implies common locations for the carriers of each DIBs on these sightline.

\begin{table}
\centering
\caption{DIB–DIB correlation coefficients.}
\label{tab:DIB Correlation Coefficients}
\begin{tabular}{@{}cccc@{}}
\toprule
DIB    & 13177 & 14680 & 15272 \\ \midrule
13177 & -      & 0.92$\pm$0.02   & 0.95$\pm$0.02   \\
14680 & 0.92$\pm$0.02   & -      & 0.95$\pm$0.01   \\
15272 & 0.95$\pm$0.02   & 0.95$\pm$0.01   & -      \\ \bottomrule
\end{tabular}
\end{table}

\subsection{Profile and properties of the 14680\,Å DIB}
\label{sec:DIB_shape}

\begin{figure}
	\includegraphics[width=\columnwidth]{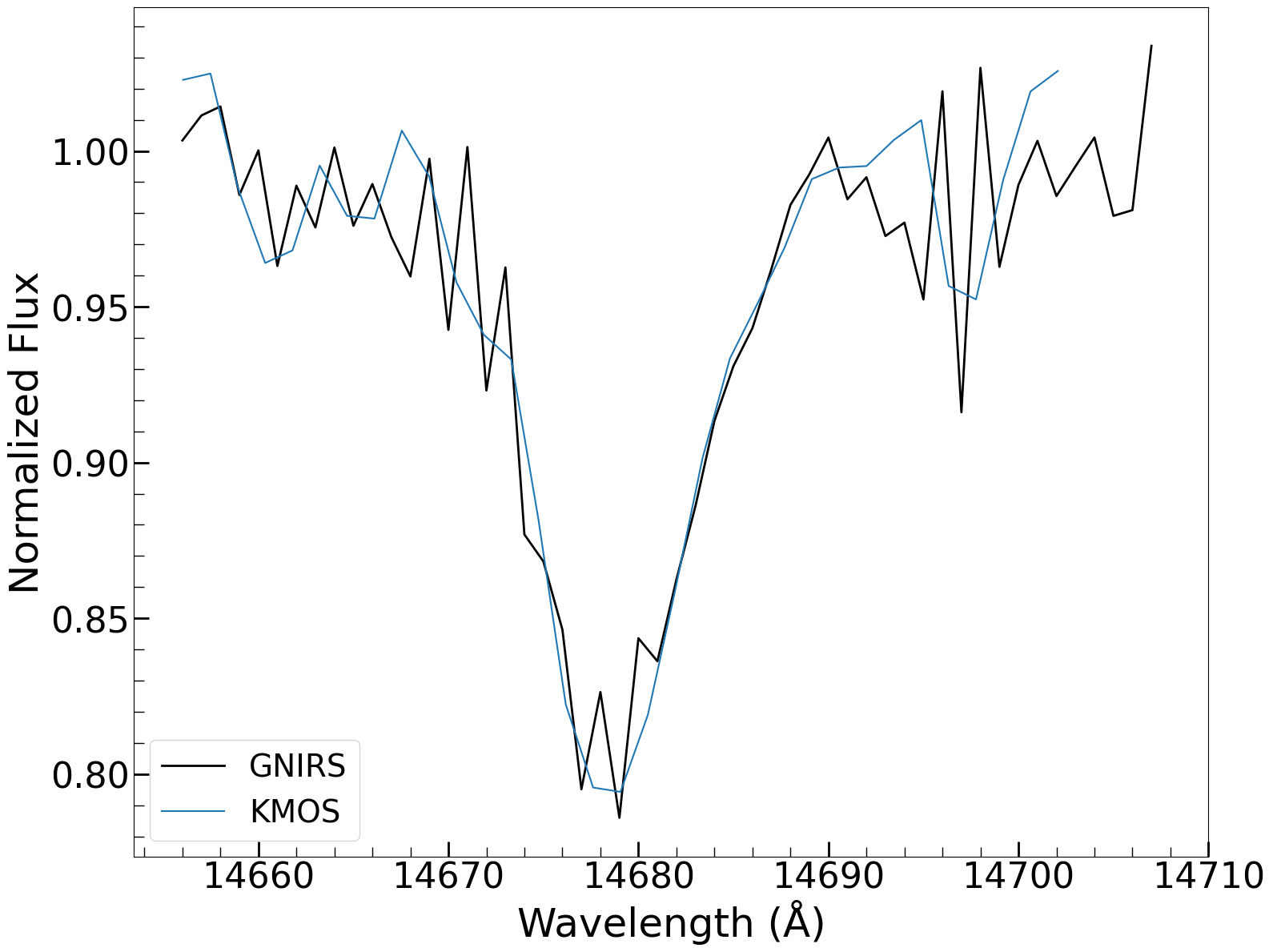}
    \caption{Comparison of $\lambda$14680\,Å DIB profile from qF362 observed using VLT-KMOS and GEMINI-GNIRS. The consistent shape of the profiles across both sources underscores the negligible influence of instrument resolution on the measured EW of the DIBs.}
    \label{fig:DIB_profile_GC}
\end{figure}

\begin{figure}
	\includegraphics[width=\columnwidth]{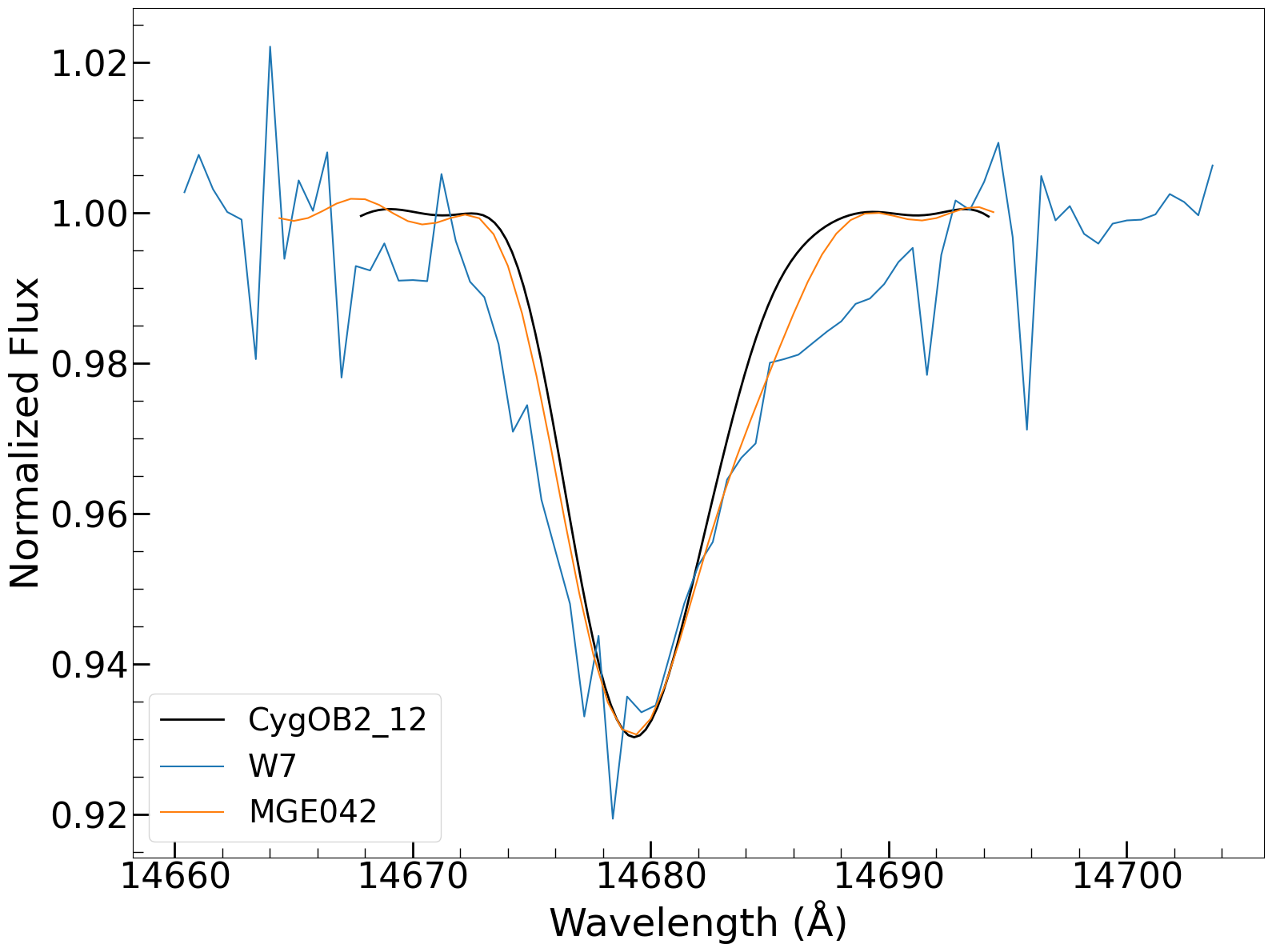}
    \caption{$\lambda$14680\,Å DIB spectra observed in objects with varying reddening. The profiles have been vertically linearly rescaled and the central wavelengths shifted to enable direct comparison. The reference profile is based on CygOB2 12, serving as a representative baseline.}
    \label{fig:1.46_DIB_profile}
\end{figure}

\begin{figure}
	\includegraphics[width=\columnwidth]{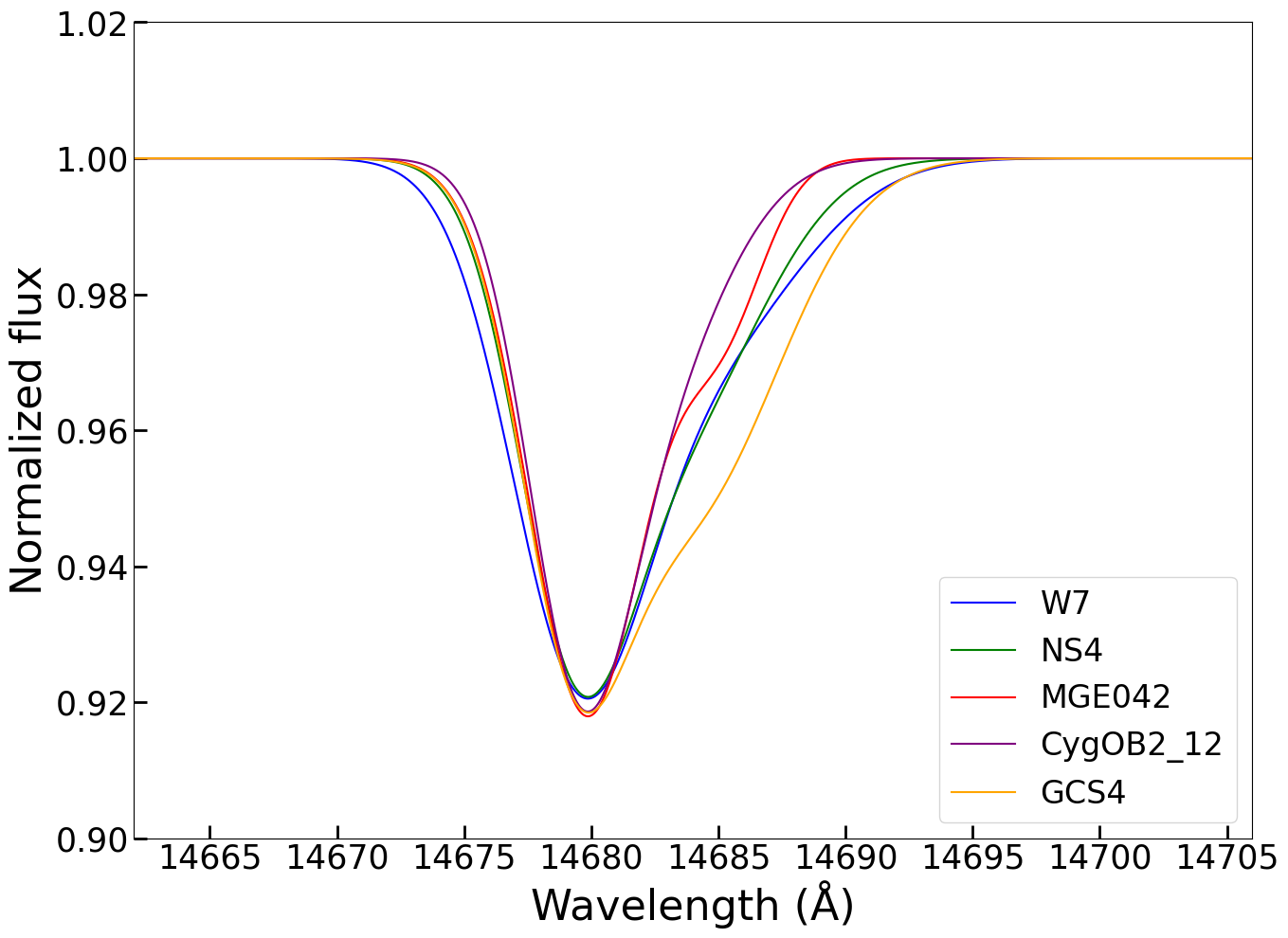}
    \caption{$\lambda$14680\,Å DIB Gaussian fits for different sightlines. The spectra have been normalized to its minimal value, and their wavelengths shifted to align at the minimum of the fit.}
    \label{fig:1_46_componentes}
\end{figure}

In Figs. \ref{fig:Gaussian_fits}, \ref{fig:DIB_profile_GC} and \ref{fig:1.46_DIB_profile} it can be seen that the profile of the $\lambda$14680\,Å DIB is asymmetric, with an extended red wing, and cannot be fit well by a single Gaussian.
This type of asymmetry is similar to that observed in most optical \citep{Krelowski_1997_profiles} and some NIR DIBs, including the infrared ones located at $\lambda$15617\,Å and $\lambda$15673\,Å \citep{elyajouri2017near}. 
Fig. \ref{fig:DIB_profile_GC} shows that although the spectrometers that we used have widely different resolutions, all of them easily resolved this DIB and hence indicate that the asymmetry is ubiquitous.
The different behaviours of the two Gaussian components needed to fit the profile on different lines of sight suggests that two distinct carriers are responsible for this DIB.
To test this, we investigate possible variations in line shape by normalizing the $\lambda$14680\,Å DIB to its minima for different sightlines and reddening.
Fig. \ref{fig:1_46_componentes} shows, that, indeed the shape of the red component of the $\lambda$14680\,Å DIB seems to be decoupled from the blue one, strengthening the argument for a second carrier. 

The same asymmetric profile of the $\lambda$14680\,Å DIB was also observed towards BD+40 4220 and HD229059 by \citet{galazutdinov2017infrared}. It therefore seems unlikely that the observed enhancement of the red wing is purely attributable to velocity structure.
While the profile of the $\lambda$14680\,Å DIB is invariant on most lines of sight, stars in the Wd1 region show slightly broader profiles which persist after convolution to a common resolution, as shown in blue in Fig. \ref{fig:1.46_DIB_profile}, and as can be seen in the Wd1 spectra from both VLT-KMOS and VLT-Xshooter. 
The broadening towards Wd1 may result from the presence of multiple foreground clouds with different radial velocities.
To determine whether the enhanced EW of this DIB in Wd1 is due to favourable conditions in a single molecular cloud or to the cumulative effects of several clouds, and to investigate possible substructures within this band, high spectral resolution studies of objects at different distances along this line of sight are required.
Additionally, it is important to consider that this broadening may also be influenced by variations in the excitation conditions of the molecules responsible for the DIB. As proposed by \citet{MacIsaac_2022} in optical DIBs, changes in the kinetic temperature and other environmental factors could similarly impact the absorption characteristics of NIR DIBs, potentially accounting for the observed variations without the need for invoking multiple carriers.

\subsection{Broadening of NIR DIBs profiles towards the Galactic Centre}
\label{sec:GC_broadening}

\begin{figure}
	\includegraphics[width=\columnwidth]{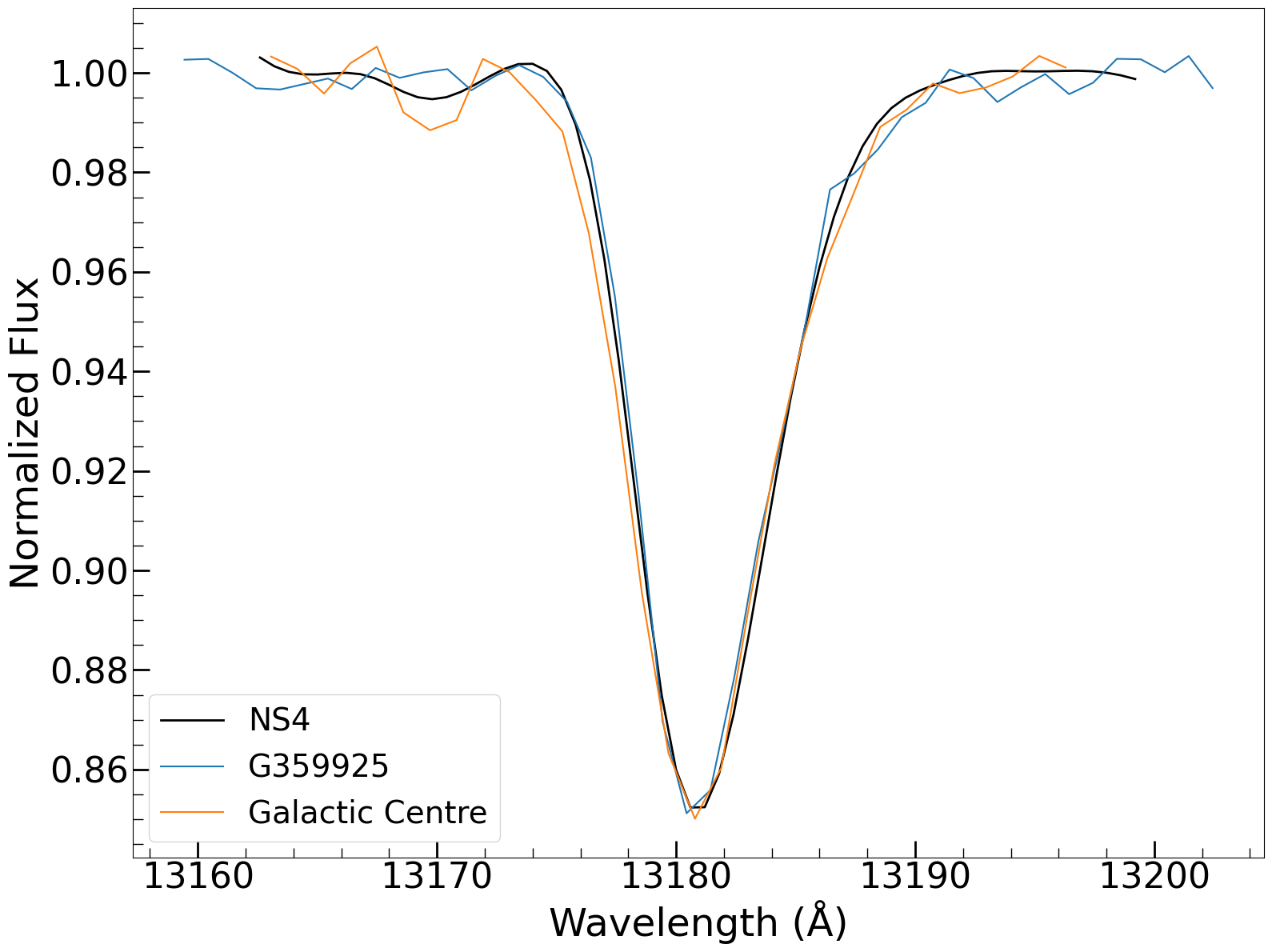}
    \caption{$\lambda$13177\,Å DIB spectra of sources on the GC (orange trace). The NS4 spectrum has been convolved to match the resolution of the spectra of the GC stars. The profiles have been vertically linearly rescaled and the central wavelengths shifted to enable direct comparison.}
    \label{fig:1_31_GC}
\end{figure}

\begin{figure}
	\includegraphics[width=\columnwidth]{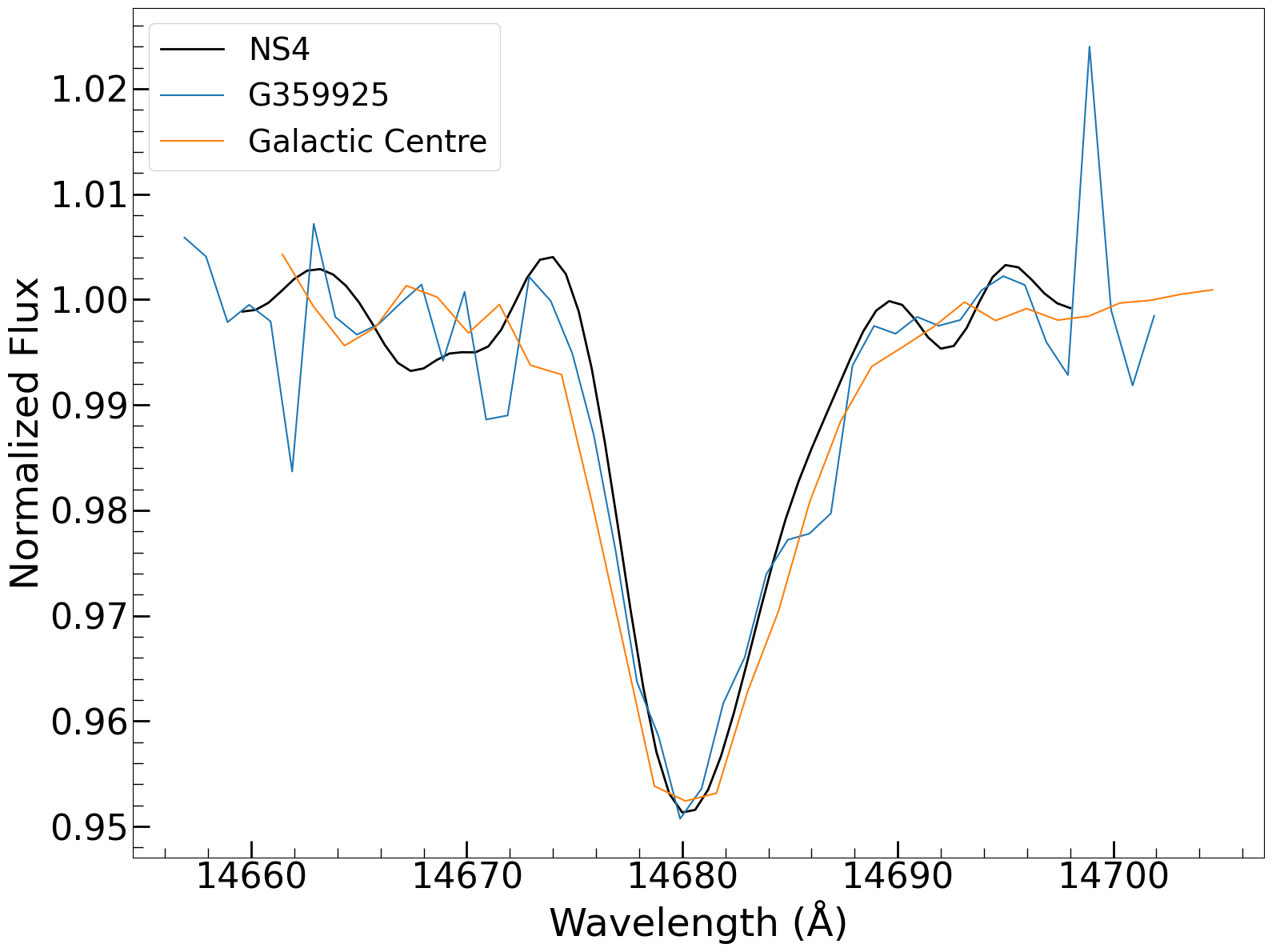}
    \caption{Analogue to Fig. \ref{fig:1_31_GC}, but for the profile of $\lambda$14680\,Å DIB. See Fig. \ref{fig:1_31_GC} caption for labels.}
    \label{fig:1_46_GC}
\end{figure}

\begin{figure}
	\includegraphics[width=\columnwidth]{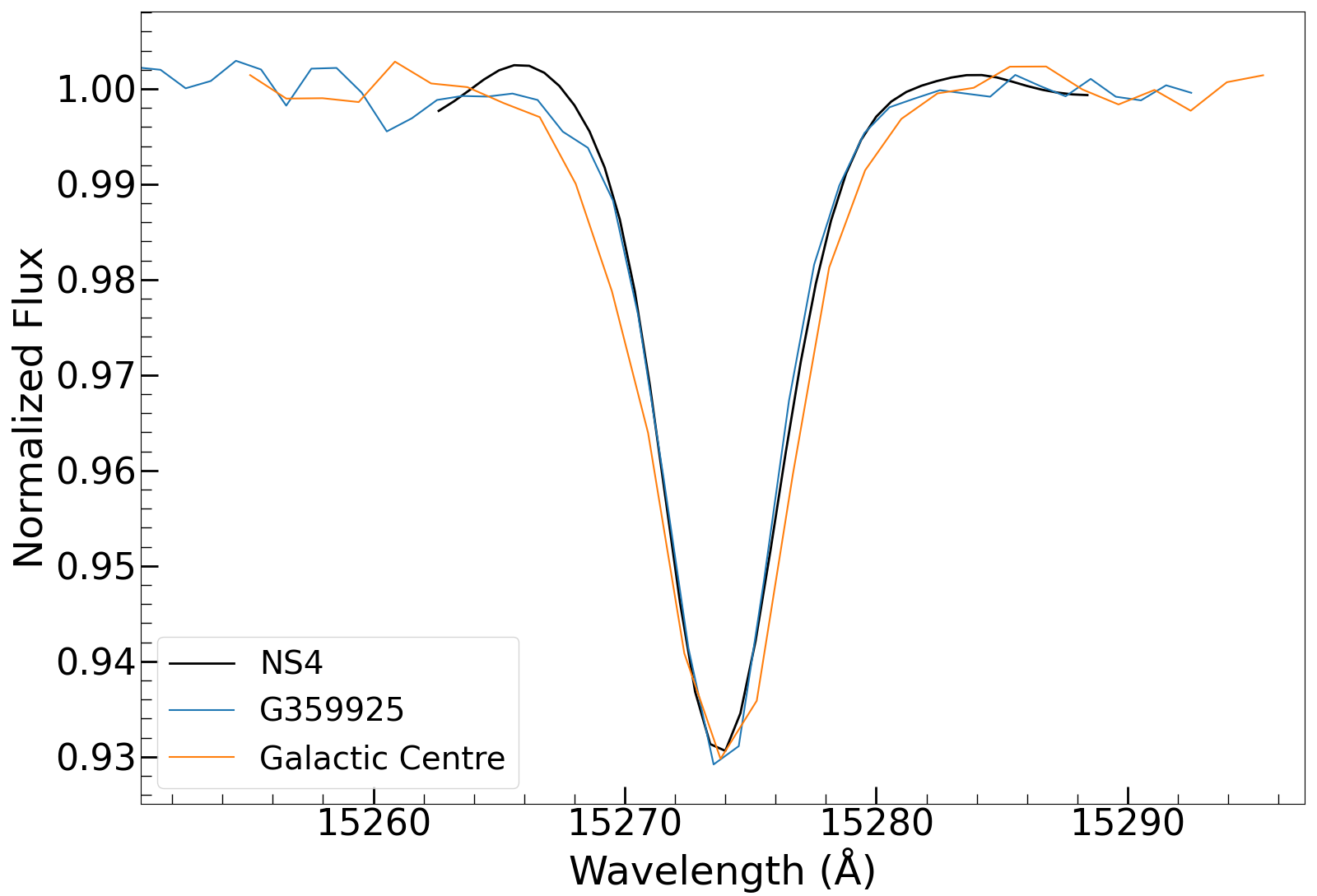}
    \caption{Analogue to Fig. \ref{fig:1_31_GC}, but for the profile of $\lambda$15272\,Å DIB. See Fig. \ref{fig:1_31_GC} caption for labels.}
    \label{fig:1_52_GC}
\end{figure}

In this section, we examine the profiles of the $\lambda$13177\,Å, $\lambda$14680\,Å and $\lambda$15272\,Å DIBs towards stars in the GC. 
The GC spectrum we use is the average of all the spectra of stars within 0.17 degrees of the Sgr A*, the supermassive black hole at the centre of the Milky Way.
As described earlier, interstellar material in dense and diffuse clouds is not uniformly distributed along sightlines but is expected to be more concentrated in the spiral arms. 
It is well known that three spiral arms cross sightlines to the GC (see e.g. \citet{Oka_2019} and references therein), each characterised by a distinct radial velocity: the $\sim$3 kpc arm \citep{Rougoor_1960} shows velocities around -50 km/s, the 4.5 kpc arm \citep{Menon_1970} around -30 km/s, and the Local Arm near 0 km/s.

In our sample, star NS4, at a distance of 4 kpc from the Sun, has its sightline only six degrees away from the sightline to Sgr A*, which is 8 kpc distant.
Comparing the DIBs profiles in the line of sight of NS4 with those towards sources in the GC may help to reveal how the different contributions of interstellar material affects the profile of the DIBs in this line of sight. 
G359925 at a distance of 5.8 kpc \citep{Bailer_Jones_21} from the Sun and on the same line of sight as the stars located in the GC, serves as an intermediate benchmark, allowing differences of the profiles towards the GC to be followed.

The line of sight to NS4, crosses the material from the -30 km/s spiral arm, while the sighline towards G359925 is additionally influenced by the -50 km/s spiral arm.
Figs. \ref{fig:1_31_GC}, \ref{fig:1_46_GC} \& \ref{fig:1_52_GC} show that the profiles of the DIBs in the lines of sight to these two stars are practically identical.
Notably, there is a minor distinction: the $\lambda$14680\,Å DIB profile of G359925 star exhibits a slightly wider blue wing compared to NS4.
The difference in the radial velocity of the interstellar material affecting these lines of sight, together with the irregular distribution of the material containing the carriers of these DIBs, could cause the slight dissimilarity between the two profiles.

The mean profile of the stars in the GC shows broadening of $\sim$2\,Å in the blue wings of the different DIBs.
This suggests the presence of carriers in the interstellar material closer to the GC with higher negative radial velocities than in the spiral arms as the cause of this broadening. 
DIBs located in the 3 kpc arm, which has a more highly blueshifted radial velocity than the 4.5 kpc and local arms (by $\sim$20 and $\sim$50 km/s, respectively), are a possible cause of the extended blue wing. 
In addition, observations of diffuse molecular gas in the CMZ \citep{Oka_2020} have reveal the presence of blue-shifted diffuse gas with velocities up to -150 km/s.
However, the profiles of the DIBs toward stars in the GC are narrower than would be expected from such high radial velocities.
The hostile conditions present in the CMZ, characterised by high cosmic ray densities \citep{Oka_2019}, could inhibit the formation of the carriers of these DIBs in the CMZ, and thereby explain the observed smaller broadening of the DIBs.

\subsection{Detection of a new Diffuse Interstellar Band}
\label{sec:New_DIBs}

\begin{figure}
	\includegraphics[width=\columnwidth]{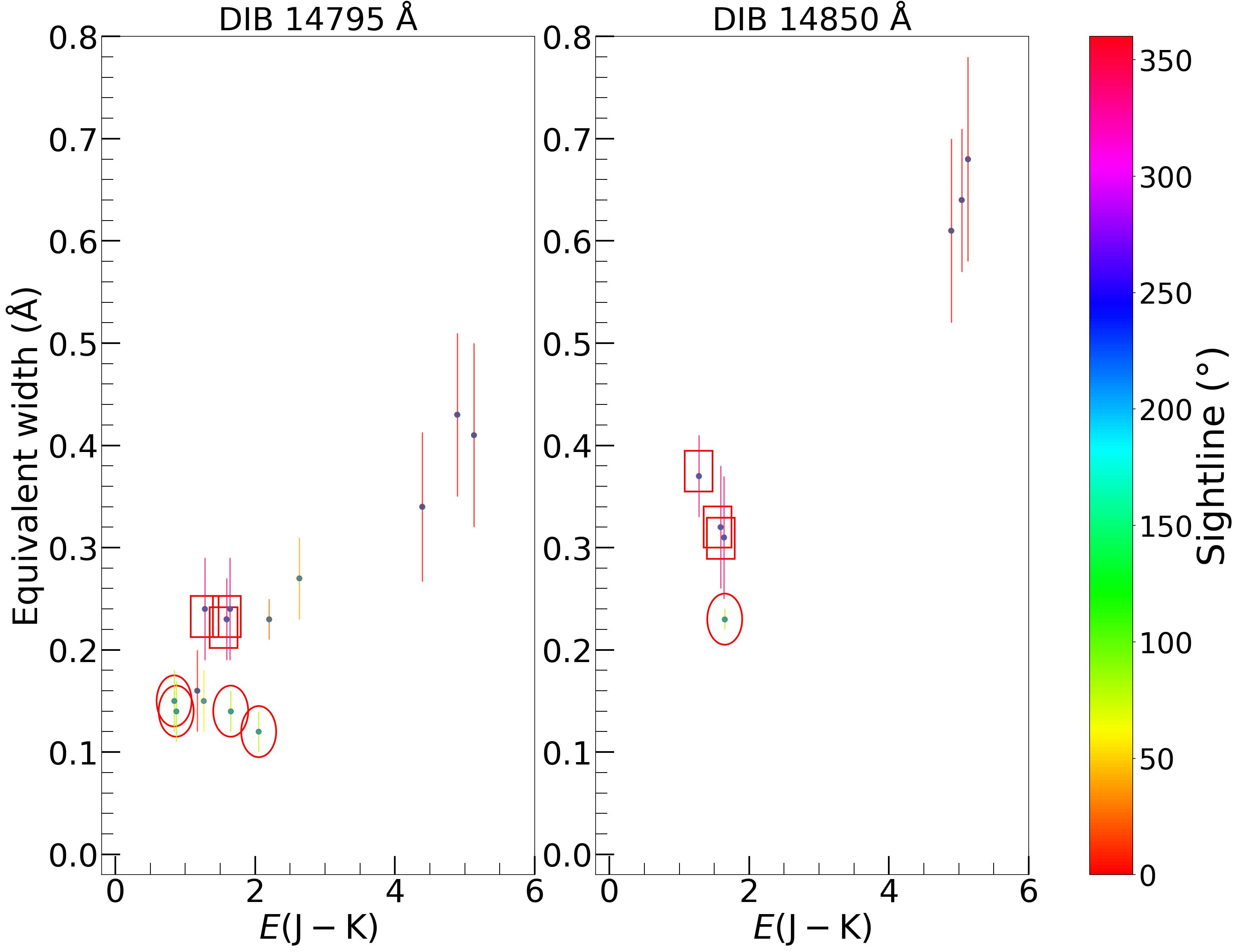}
    \caption{Equivalent width in\,Å vs. $E(J-K)$ index for the objects where the new $\lambda$14795\,Å and $\lambda$14850\,Å DIBs have been detected. The data points of the CygOB2 association are marked as red circles, while Wd1 cluster members are marked with red squares.}
    \label{fig:New_DIBs_JK}
\end{figure}

The spectra presented here revealed the presence of two faint DIBs in wavelength intervals  heavily contaminated by telluric lines at the short wavelength edge of the H-band.
One of these DIBs, located at $\lambda$14850\,Å, was first reported in two sources by \citet{Najarro_GC_2017}, while the other, at $\lambda$14795\,Å, is a new detection.
The stars in which these were detected are not expected to have strong atomic lines at these wavelengths, which reduces the possibility of confusion with other spectral features.
Neither DIB was detected towards stars with very low reddening, which is consistent both with there not being stellar lines in these wavelength intervals and with the expected behaviour of DIBs based on the nearly linear correlation between DIB strength and reddening.
The differences in their profiles, one narrow and the other showing a typical broad DIB profile (see Figs. \ref{fig:Examples_DIB_14795} \& \ref{fig:Examples_DIB_14850}), suggest different carriers for these bands. 
Notably, the DIB at $\lambda$14850\,Å has fewer measurements compared to the DIB at $\lambda$14795\,Å, which can be attributed to its broader and shallower shape, and the higher S/N required for its detection. 

In the sources toward which these bands were detected, we performed Gaussian fits following the same method described in Sect.\ref{sec:Results}. 
In Fig. \ref{fig:New_DIBs_JK}, we include the EWs measured in those sources where a detection threshold of 5$\sigma$ is exceeded (see Table~\ref{tab:New_DIBs_Values}).
The dependencies of EW on $E(J-K)$ of these DIBs qualitatively follow the trends seen in fig.\ref{fig:DIBs_ew_reddening} for the $\lambda$13177\,Å, $\lambda$14680\,Å and $\lambda$15272\,Å DIBs. 
This strengthens the conclusion that these features correspond to DIBs.

The EWs of the $\lambda$14795\,Å DIB toward CygOB2 association members (red circles in Fig. \ref{fig:New_DIBs_JK}) show a nearly identical value despite the differences in reddening. 
This behaviour is similar to those shown by $\lambda$14680\,Å and $\lambda$15272\,Å DIBs, reinforcing the idea that the carriers of these DIBs are not located within the dust responsible for the local obscuration of each star.
In the subset of stars within Wd1 cluster (highlighted with red squares in Fig. \ref{fig:New_DIBs_JK}) where these DIBs have been detected, their EWs slightly exceed the anticipated regression line. 
This trend is line with the behaviour seen for the $\lambda$15272\,Å and, most notably, the $\lambda$14680\,Å DIB towards this line of sight.
This recurring phenomenon supports the idea that molecular clouds towards Wd1 present more favourable conditions for the existence of the carriers that are responsible for these DIBs.


\clearpage
\onecolumn
\begin{landscape}
\footnotesize
\setlength\tabcolsep{6pt} 
\setlength{\LTpost}{0pt} 
\begin{longtable}{lccccccccc}
\caption{List of objects and EWs values for the $\lambda$14795 Å and the $\lambda$14850 Å NIR DIBs. Columns with no entries did not meet the 5$\sigma$ threshold.} \label{tab:New_DIBs_Values} \\
\hline 
Star \textit{(Alt. ID)} & Instrum & \text{Sightline \text{(º)}\footnotemark[1]} & Sp type & J & K & Ref.\footnotemark[2] & $E(J-K)$\footnotemark[3] & \text{$W_{\lambda14795}$}\text{(Å)} & \text{$W_{\lambda14850}$}\text{(Å)}  \\

\hline
\hline
\endfirsthead

\hline
\multicolumn{10}{r}{{c}} \\
\endfoot

\hline
\vspace{-20mm} 
\endlastfoot

TIC 195381536 (\textit{J2039}) & GIANO-B & 81.82 & B0I & 7.34 & 5.82 & 1 & 1.65 & 0.14 $\pm$ 0.02 & 0.23 $\pm$ 0.01  \\
TIC 175996914 (\textit{GAL026}) & GIANO-B & 26.46 & LBV & 7.99 & 5.61 & 1 & 2.20 & 0.23 $\pm$ 0.02 & - \\
CygOB2 7 & GIANO-B & 80.24 & O3If* & 7.25 & 6.61 & 1 & 0.85 & 0.15 $\pm$ 0.02 & -  \\
CygOB2 11 & GIANO-B & 80.56 & O5.5Ifc & 6.65 & 5.99 & 1 & 0.86 & 0.14 $\pm$ 0.02 & -  \\
CygOB2 12 & GIANO-B & 80.10 & B3-4Ia+ & 4.66 & 2.7 & 1 & 2.05 & 0.12 $\pm$ 0.02 & -  \\
WR 105 (\textit{NS4})  & GIANO-B & 6.52 & WN9h & 7.03 & 5.73 & 1 & 1.17 & 0.16 $\pm$ 0.03 & -  \\
TIC 196042957 (\textit{MGE042}) & GIANO-B & 42.07 & LBV & 9.66 & 6.85 & 1 & 2.63 & 0.27 $\pm$ 0.04 & -  \\
IRAS 19425+2411 (\textit{MN112}) & GIANO-B & 60.45 & LBV & 8.86 & 7.42 & 1 & 1.26 & 0.15 $\pm$ 0.03 & - \\
\lbrack DWC2011\rbrack\ 103 & Xshooter & 0.05 & OI & 13.33 & 9.02 & 1 & 4.34 & 0.36 $\pm$ 0.06 & -  \\
Cl* Westerlund1 W7 & Xshooter & 339.56 & B5 Ia+ & 6.94 & 5.41 & 2 & 1.59 & 0.23 $\pm$ 0.04 & 0.32 $\pm$ 0.06  \\
Cl* Westerlund1 W33  & Xshooter & 339.55 & B5 Ia+ & 6.88 & 5.29 & 2 & 1.64 & 0.24 $\pm$ 0.04 & 0.31 $\pm$ 0.06  \\
Cl* Westerlund1 W57a & Xshooter & 339.53 & B4 Ia & 8.13 & 6.93 & 2 & 1.28 & 0.24 $\pm$ 0.04 & 0.37 $\pm$ 0.04  \\
qF 362 & GNIRS & 0.17 & LBV & 12.31 & 7.09 & 1 & 5.04 & -  & 0.64 $\pm$ 0.07  \\
GCS 4 \footnotemark[4] & GNIRS & 0.16 & WC8/9d+OB & 14.53 & 7.23 & 3 & 4.89 & 0.43 $\pm$ 0.08 & 0.61 $\pm$ 0.09  \\
\lbrack DWC2011\rbrack\ 98 (\textit{G0.070}) & GNIRS & 0.06 & O4-6If+ & 14.79 & 9.86 & 4 & 5.13 & 0.45 $\pm$ 0.08 & 0.68 $\pm$ 0.11  \\
\end{longtable}

\begin{minipage}{0.95\linewidth}
\vspace{12mm}
\renewcommand{\thempfootnote}{\arabic{mpfootnote}}
\footnotetext[1]{Angle relative to that toward Sgr A* in the GC.}
\footnotetext[2]{Reference for photometry: (1) - \citet{Cutri_2mass_2003} (2) - \citet{Gennaro_2017}, (3) - \citet{Clark_2018}, (4) - \citet{Mauerhan_2010}.}
\footnotetext[3]{The intrinsic colours of the stars are derived from specific sources: OB types are derived from \citet{Wegner_1994}, WR stars are based on \citet{Crowther_2006}. LBVs are approximated to the intrinsic colour of WN11h stars, reflecting their spectral and luminosity characteristics.}
\footnotetext[4]{Sources where the dust contribution to the $E(J-K)$ colour index has been removed following \citet{Najarro_GC_2017}.}
\end{minipage}

\end{landscape}
\clearpage
\twocolumn

\section{Conclusions}
\label{sec:Conclusions}

We have analyzed the DIBs in the 13000-15300\,Å spectra of a large sample of stars obscured by interstellar material, most of which is in the diffuse interstellar medium. The data cover a wider range of extinctions (up to Av $\sim$ 30) than previous work. 
We have concentrated our analysis on three prominent DIBs at $\lambda$13177\,Å, $\lambda$14680\,Å and $\lambda$15272\,Å.
The key findings of our study are:

\begin{itemize}
\item The equivalent widths of the $\lambda$13177\,Å, $\lambda$14680\,Å and $\lambda$15272\,Å DIBs correlate well with reddening in all observed environments, confirming previous results at lower extinction and reinforcing their interstellar origin. However, while the DIB-DIB correlation is robust, the high S/N of our measurements reveals that the pair correlations are not tight enough for us to conclude that any pair of these three DIBs arises from the same carrier. 
\item The strength of the $\lambda$13177\,Å DIB exceeds that of the $\lambda$14680\,Å and $\lambda$15272\,Å DIBs on all lines of sight studied consistent with previous studies, indicating a higher abundance of its carrier and/or a stronger transition strength than those of the other two bands.

\item The $\lambda$14680\,Å DIB can be accurately modelled by the superposition of two Gaussians, the second required to account for its extended red wing. 
This consistent profile structure across numerous sightlines suggest that the DIB profile variations are intrinsic, implying a connection to the band formation process rather than velocity-related structural changes. The different behaviours on different lines of sight of the two Gaussian components of the fit for this DIB suggests that the absorption could be a blend of two distinct DIBs. However, the study by \citet{MacIsaac_2022} shows that difference in kinetic temperature in different DIB forming regions also can account for variations in the absorption profiles of several optical DIBs, an effect which could apply to NIR DIBs. 

\item We identify one new DIB at $\lambda$14795\,Å. 

\item The EWs of the $\lambda$14680\,Å, $\lambda$14795\,Å and $\lambda$15272\,Å DIBs are roughly constant for stars in the CygOB2 association, implying that the carriers are not formed efficiently within the OB association. 
Conversely, the EW of the $\lambda$13177\,Å DIB varies significantly from star to star in CygOB2, indicating a different origin for its carrier and suggesting that its carrier is more resistant to the intense radiation fields present in CygOB2 than those of the other bands.

\item The EWs of the DIBs on sightlines toward Wd1 are stronger than those on other lines of sight with similar extinctions. The difference is largest for the $\lambda$14680\,Å DIB. This suggests that molecular clouds along this line of sight are more conducive to the formation of carriers of this DIB.

\item The near constancy of the ratios of EWs between the three main studied DIBs favours similar environmental dependencies of their abundances.

\item The profiles of the DIBs toward the stars NS4 and G359925, whose lines of sight are in the direction of the GC, are slightly broadened compared to those on shorter sightlines, suggesting that their profiles are affected by gas in the intervening 4.5 kpc spiral arm. 
The DIBs profiles toward stars in the GC show a larger broadening on their short wavelength edges, which may be due to DIBs carriers located in the inner 3 kpc arm. The absence of even more highly blueshifted wings indicates that the DIBs carriers are not located in the rapidly radially expanding diffuse molecular gas located in the CMZ, suggesting that the environment there is hostile to the formation of DIBs carriers.

\end{itemize}

Our results emphasize the importance of infrared spectroscopy in the exploration of highly reddened regions of space, a task now more accessible due to advances in space-based infrared astronomy. 
With state-of-the-art instruments such as the Near-Infrared Spectrograph (NIRSpec) on the JWST enabling observations of highly reddened regions of the Universe, astronomers now have access to parts of the electromagnetic spectrum that were previously inaccessible due to Earth's atmospheric telluric contamination.


\section*{Acknowledgements}

R.C. gratefully  acknowledges the financial support through the PRE2020-096167 grant provided by the Spanish MCIN.
R.C., F.N., M.G. and L.R.P. acknowledge support by grants PID2019-105552RB-C41 and PID2022-137779OB-C41 funded by MCIN/AEI/10.13039/501100011033 by "ERDF A way of making Europe". 
Based on observations collected at the European Southern Observatory under ESO programme(s) 098.D-0373(A), 0103.D-0316(A) and 109.233D.001 and data obtained from the ESO Science Archive Facility with DOI: https://doi.org/10.18727/archive/71.
This research used the facilities of the Italian Center for Astronomical Archive (IA2) operated by INAF at the Astronomical Observatory of Trieste.
This research is based in part on observations for programs GN-2011A-Q-54, GN-2011A-Q-64, GN-2012A-Q-61, GN-2013A-Q-55, and GN-2014A-Q-89) obtained at the international Gemini Observatory, a program of NSF's NOIRLab, which is managed by the Association of Universities for Research in Astronomy (AURA) under a cooperative agreement with the National Science Foundation, on behalf of the Gemini Observatory partnership: the National Science Foundation (United States), National Research Council (Canada), Agencia Nacional de Investigaci\'{o}n y Desarrollo (Chile), Ministerio de Ciencia, Tecnolog\'{i}a e Innovaci\'{o}n (Argentina), Minist\'{e}rio da Ci\^{e}ncia, Tecnologia, Inova\c{c}\~{o}es e Comunica\c{c}\~{o}es (Brazil), and Korea Astronomy and Space Science Institute (Republic of Korea). 

\section*{Data Availability}


VLT data are available in the ESO Science Archive Facility (http://archive.eso.org/wdb/wdb/adp/phase3\_main/form).
TNG data are available in the TNG Archive (http://archives.ia2.inaf.it/tng/).
Gemini data are available in the Gemini Archive (https://archive.gemini.edu/searchform).



\bibliographystyle{mnras}
\bibliography{Bibliography} 




\appendix

\section{DIBs contamination}
\label{sec:Appendix_A}

The Brackett 19 line at $\lambda$15264\,Å in stellar atmospheres is a potential contaminant of the $\lambda$15272\,Å DIB. 
This hydrogen line could have either a P-Cygni profile or be purely in emission for LBVs and is in absorption in B-super and -hypergiants.
To obtain precise measurements of the DIB in these cases, a Gaussian fitting technique was utilized for the 20th, 18th, 17th, and 16th transitions of the Brackett sequence. 
This approach yielded the EWs for these Bracket lines and enabled an accurate EW of Br19 to be determined through interpolation.

\begin{figure}
	\includegraphics[width=\columnwidth]{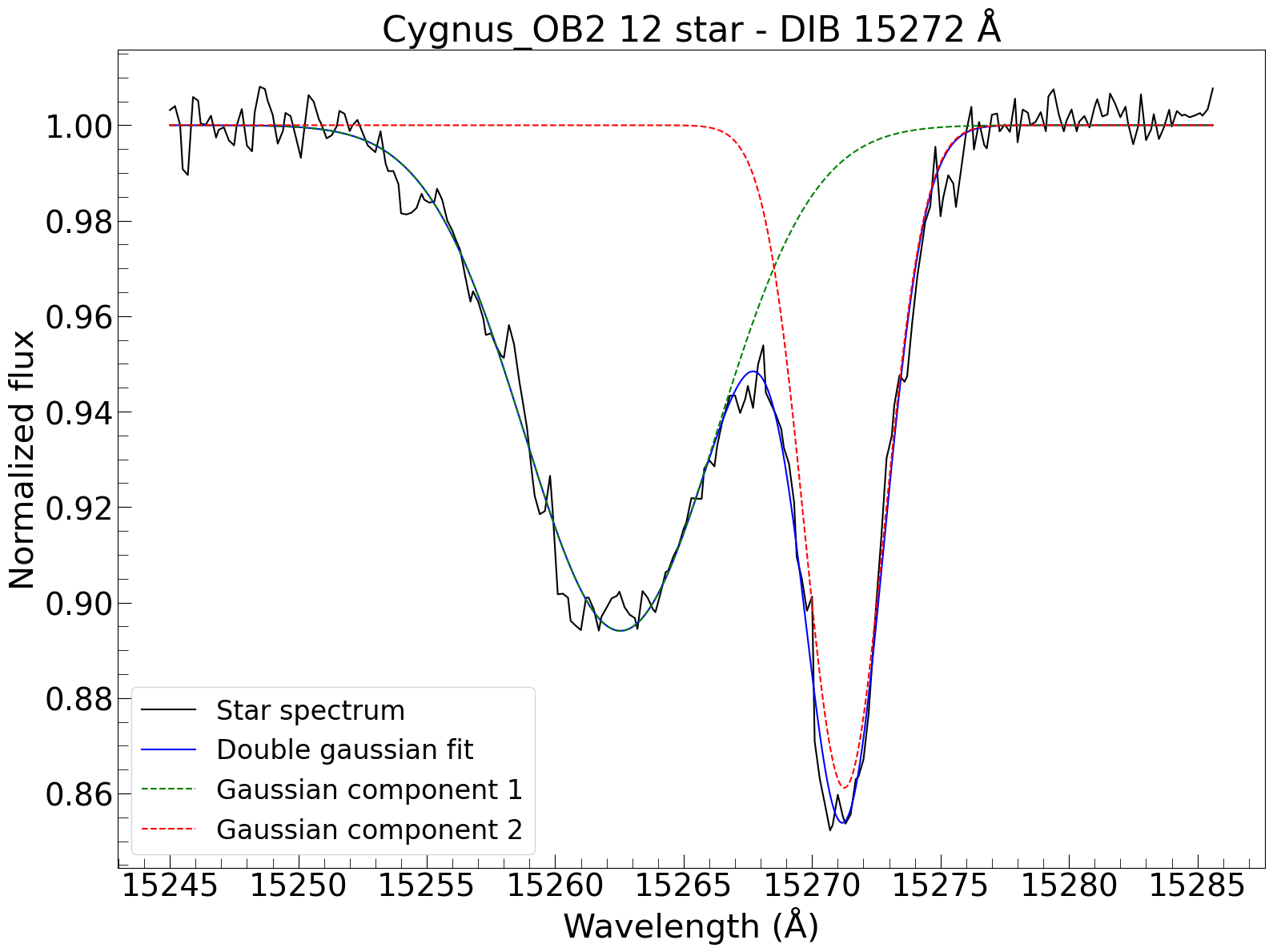}
    \caption{Double Gaussian fitting of the $\lambda$15272\,Å DIB (in green) and HI 19 line (in red) is shown, illustrating the separation of the spectral absorption features. This fitting process is crucial in precisely quantifying the equivalent width of the $\lambda$15272\,Å DIB for our analysis.}
    \label{fig:15277_contamination}
\end{figure}

To test the validity of the above method we also performed a double Gaussian fitting, modelling both the DIB and hydrogen lines as illustrated in Fig. \ref{fig:15277_contamination}. 
The EWs so obtained for the DIB were found to be in good agreement with those acquired using the first method. 

None of the stars included in this study displayed detectable lines from the Brackett hydrogen series beyond the 24th transition at $\lambda$15000\,Å. As a result, the $\lambda$14680\,Å, $\lambda$14795\,Å, and $\lambda$14850\,Å DIBs were not influenced by the Brackett lines.

\begin{figure}
	\includegraphics[width=\columnwidth]{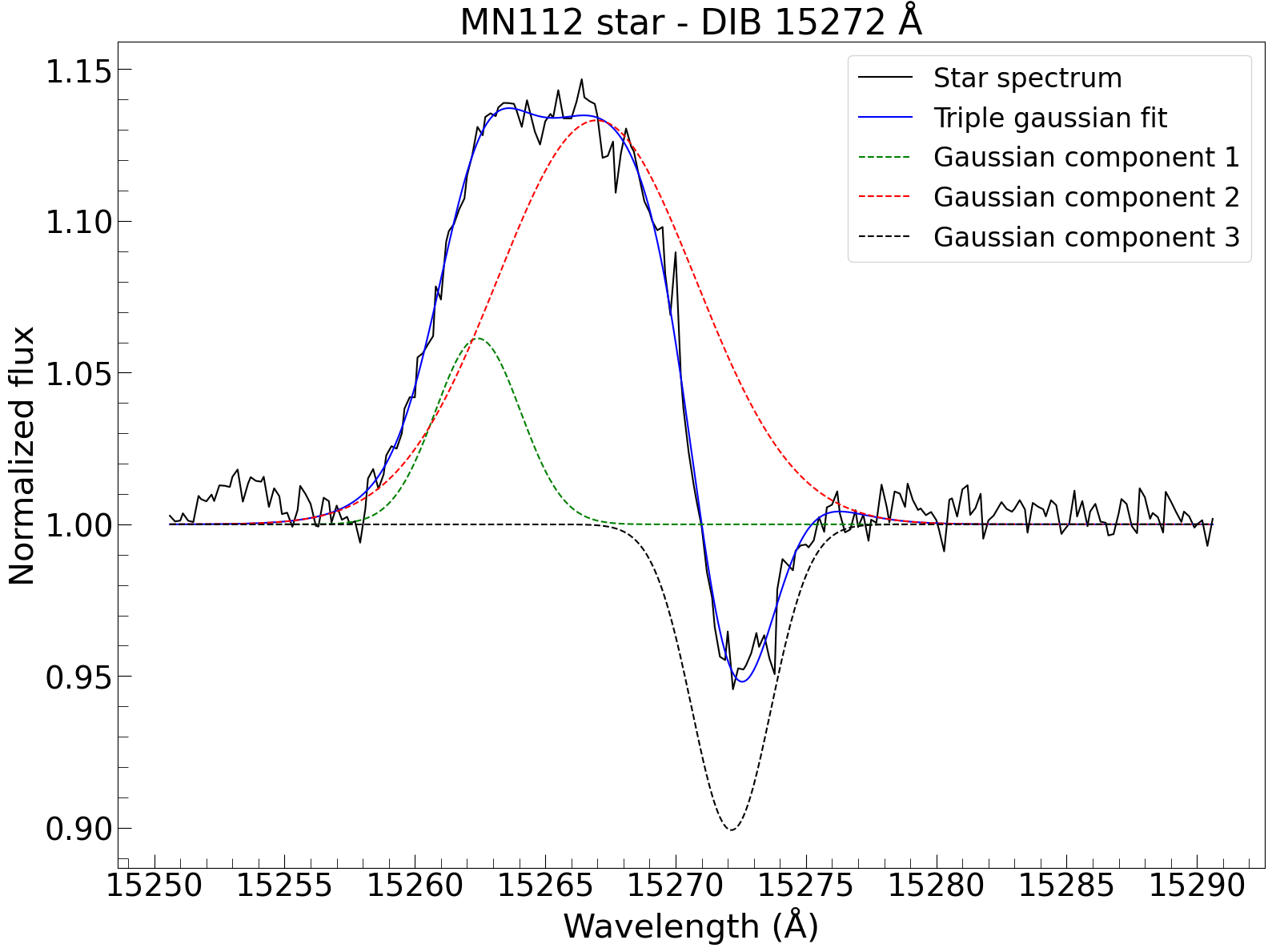}
    \caption{Triple Gaussian fitting is applied to resolve the $\lambda$15272\,Å DIB absorption feature (in black) from the adjacent HI 19 line (in red) and NII line contamination (in green).}
    \label{fig:15277_doble_contamination}
\end{figure}

In the spectra of two LBVs, this DIB is contaminated by not only the 19 Brackett line but also the NII emission line at $\lambda$15269\,Å. 
To account for this, we applied the methodology described earlier, using a triple Gaussian fitting approach to fit the $\lambda$15272\,Å band, the hydrogen line, and the nitrogen line simultaneously. 
The outcome is shown in Fig. \ref{fig:15277_doble_contamination}.
The EW of the DIB is the EW of the Gaussian used to fit the DIB's contribution to the observed profiles.


\section{Sample coordinates \& distances}
\label{sec:Appendix_B}

\onecolumn
\footnotesize
\setlength\tabcolsep{4pt} 
\setlength{\LTpost}{0pt} 
\begin{longtable}{lccccccc}
\caption{List of objects included in this paper.} \label{tab:Targets_info} \\
\hline 
{Star \textit{(Alt. ID)}} & {Instrum} & {Sightline (º)\footnotemark[1]} & {RA (J2000.0)} & {Dec (J2000.0)} & {Distance (kpc)} & Ref\footnotemark[2] & Cluster \\
& & & & & & & \\
\hline
\hline
\endfirsthead

\multicolumn{8}{c}%
{{\tablename\ \thetable{} -- \textit{continued.}}} \\
\hline
{Star \textit{(Alt. ID)}} & {Instrum} & {Sightline (º)\footnotemark[1]} & {RA (J2000.0)} & {Dec (J2000.0)} & {Distance (kpc)} & Ref\footnotemark[2] & Cluster \\
& & & & & & & \\
\hline
\hline
\endhead

\hline
\vspace{-20mm} 
\endlastfoot

    WR42e & KMOS & 291.57 & 11:14:45.51 & -61:15:00.29 & 5.75 & 1 & NGC3603 \\
    TIC 467481534 (\textit{MTT31}) & KMOS & 291.62 & 11:15:06.68 & -61:16:33.43 & 5.75 & 1 & NGC3603 \\
    \lbrack CRN2020\rbrack\ W1008 & KMOS & 339.52 & 16:46:55.40 & -45:51:54.26 & 4.23 & 2 & Wd1\\
    \lbrack CRN2020\rbrack\ W1025 & KMOS & 339.53 & 16:47:00.77 & -45:52:04.77 & 4.23 & 2 & Wd1\\
    \lbrack CRN2020\rbrack\ W1034 & KMOS & 339.53 & 16:47:02.52 & -45:51:48.42 & 4.23 & 2 & Wd1\\
    \lbrack CRN2020\rbrack\ W1043 & KMOS & 339.55 & 16:47:04.56 & -45:50:59.46 & 4.23 & 2 & Wd1\\
    \lbrack CRN2020\rbrack\ W1047 & KMOS & 339.53 & 16:47:06.11 & -45:52:32.24 & 4.23 & 2 & Wd1\\
    \lbrack CRN2020\rbrack\ W1062 & KMOS & 339.56 & 16:47:10.64 & -45:50:46.46 & 4.23 & 2 & Wd1\\
    \lbrack CRN2020\rbrack\ W1065 & KMOS & 339.58 & 16:47:11.60 & -45:49:22.47 & 4.23 & 2 & Wd1\\
    \lbrack CRN2020\rbrack\ W1067 & KMOS & 339.59 & 16:47:13.38 & -45:49:10.63 & 4.23 & 2 & Wd1\\
    Cl* Westerlund1 W15 & KMOS & 339.56 & 16:47:06.63 & -45:50:29.72 & 4.23 & 2 & Wd1\\
    Cl* Westerlund1 W18 & KMOS & 339.55 & 16:47:05.70 & -45:50:50.48 & 4.23 & 2 & Wd1\\
    Cl* Westerlund1 W24 & KMOS & 339.54 & 16:47:02.15 & -45:51:12.58 & 4.23 & 2 & Wd1\\
    Cl* Westerlund1 W38 & KMOS & 339.55 & 16:47:02.86 & -45:50:46.14 & 4.23 & 2 & Wd1\\
    Cl* Westerlund1 W43c & KMOS & 339.55 & 16:47:03.75 & -45:50:58.44 & 4.23 & 2 & Wd1\\
    Cl* Westerlund1 W61b & KMOS & 339.53 & 16:47:02.56 & -45:51:41.86 & 4.23 & 2 & Wd1\\
    LHO 1 & KMOS & 0.16 & 17:46:16.74 & -28:49:51.2 & $\sim$8 & 3 & Quintuplet \\
    LHO 89 & KMOS & 0.16 & 17:46:15.00 & -28:49:33.2 & $\sim$8 & 3 & Quintuplet \\
    qF 134 (\textit{Pistol Star}) & KMOS & 0.15 & 17:46:15.24 & -28:50:03.58 & $\sim$8 & 3 & Quintuplet \\
    qF 274 & KMOS & 0.17 & 17:46:17.53 & -28:49:29.00 & $\sim$8 & 3 & Quintuplet \\
    qF 406 & KMOS & 0.17 & 17:46:13.85  & -28:48:50.40 & $\sim$8 & 3 & Quintuplet \\
    WR102e & KMOS & 0.15 & 17:46:14.81 & -28:50:00.6 & $\sim$8 & 3 & Quintuplet \\
    qF 256 & GNIRS & 0.16 & 17:46:16.54 & -28:49:32.00 & $\sim$8 & 3 & Quintuplet \\
    qF 320 & GNIRS & 0.16 & 17:46:14.05 & -28:49:16.5 & $\sim$8 & 3 & Quintuplet \\
    qF 362 & GNIRS & 0.17 & 17:46:17.98 & -28:49:03.46 & $\sim$8 & 3 & Quintuplet \\
    GCS 4 & GNIRS & 0.16 & 17:46:15.86 & -28:49:45.63 & $\sim$8 & 3 & Quintuplet \\
    \lbrack DWC2011\rbrack\ 98 (\textit{G0.070}) & GNIRS & 0.06 & 17:45:41.27 & -28:51:47.70 & $\sim$8 & 4 & Isolated \\
    \lbrack DWC2011\rbrack\ 23 (\textit{G0.071}) & GNIRS & 0.07 & 17:46:10.01 & -28:55:32.40 & $\sim$8 & 4 & Isolated \\
    \lbrack DWC2011\rbrack\ 36 & GNIRS & 359.97 & 17:45:31.50 & -28:57:16.8 & $\sim$8 & 4 & Isolated \\
    TIC 126154878 & GNIRS & 0.53 & 17:47:12.25 & -28:31:21.6 & $\sim$8 & 4 & Isolated \\
    WR102ka & GNIRS & 0 & 17:46:18.12 & -29:01:36.6 & $\sim$8 & 4 & Isolated \\
    \lbrack DWC2011\rbrack\ 38 (\textit{G359925}) & GNIRS & 359.92 & 17:45:37.98 & -29:01:34.49 & 5.8 & 1 & Isolated \\
    TIC 195381536 (\textit{J2039}) & GIANO-B & 81.82 & 20:39:53.58 & +42:22:50.62 & 1.76 & 1 & CygOB2 \\
    CygOB2 7 & GIANO-B & 80.24 & 20:33:14.10 & +41:20:21.90 & 1.71 & 1 & CygOB2 \\
    CygOB2 11 & GIANO-B & 80.56 & 20:34:08.51 & +41:36:59.40 & 1.78 & 1 & CygOB2 \\
    CygOB2 12 & GIANO-B & 80.10 & 20:32:40.95 & +41:14:29.27 & 1.69 & 1 & CygOB2 \\
    TIC 109609771 & GIANO-B & 12.79 & 18:13:20.98 & -17:49:47.00 & 6.43 & 1 & Isolated \\
    TIC 175996914 (\textit{GAL026}) & GIANO-B & 26.46 & 18:39:32.24 & -05:44:20.18 & 6.02 & 1 & Isolated \\
    WR 105  (\textit{NS4}) & GIANO-B & 6.52 & 18:02:23.45 & -23:34:37.49 & 4.06 & 5 & Isolated \\
    TIC 196042957  (\textit{MGE042}) & GIANO-B & 42.07 & 19:06:24.56 & +08:22:01.43 & 4.25 & 5 & Isolated \\
    HD190603 & GIANO-B & 69.40 & 20:04:36.17 & +32:13:06.95 & 1.91 & 5 & Isolated \\
    IRAS 20298+4011 (\textit{G79}) & GIANO-B & 79.28 & 20:31:42.28 & +40:21:59.07 & 1.74 & 1 & Isolated \\
    IRAS 19425+2411 (\textit{MN112}) & GIANO-B & 60.45 & 19:44:37.59 & +24:19:05.74 & 6.93 & 6 & Isolated \\
    \lbrack DWC2011\rbrack\ 103 & Xshooter & 0.05 & 17:46:01.65 & -28:55:15.3 & $\sim$8 & 4 & Isolated \\
    \lbrack DWC2011\rbrack\ 145 & Xshooter & 359.73 & 17:45:16.90 & -29:12:17.96 & $\sim$8 & 4 & Isolated \\
    TIC 116433010  (\textit{B164}) & Xshooter & 15.07 & 18:20:30.83 & -16:10:07.52 & 1.65 & 5 & Isolated \\
    TIC 116432629  & Xshooter & 15.07 & 18:20:24.38 & -16:08:43.36 & 2.5 & 1 & Isolated \\
    HD80077  & Xshooter & 271.62 & 09:15:54.78 & -49:58:24.57 & 2.63 & 5 & Isolated \\
    Cl* Westerlund1 W7  & Xshooter & 339.56 & 16:47:03.62 & -45:50:14.36 & 4.23 & 2 & Wd1 \\
    Cl* Westerlund1 W33  & Xshooter & 339.55 & 16:47:04.12 & -45:50:48.48 & 4.23 & 2 & Wd1 \\
    Cl* Westerlund1 W57a  & Xshooter & 339.56 & 16:47:01.36 & -45:51:45.80 & 4.23 & 2 & Wd1 \\
    HD10747  & Xshooter & 298.86 & 01:41:17.90 & -75:39:49.11 & 0.74 & 1 & Isolated \\
    HD48434  & Xshooter & 208.54 & 06:43:38.64 & +03:55:57.10 & 0.75 & 1 & Isolated \\
    HD154445  & Xshooter & 19.29 & 17:05:32.25 & -00:53:31.43 & 0.24 & 5 & Isolated \\
    HD203532  & Xshooter & 309.45 & 21:33:54.57 & -82:40:59.10 & 0.29 & 1 & Isolated \\
    HD209008  & Xshooter & 65.80 & 22:00:07.92 & +06:43:02.81 & 0.42 & 1 & Isolated \\

\end{longtable}

\begin{minipage}{0.95\linewidth}
\vspace{12mm}
\renewcommand{\thempfootnote}{\arabic{mpfootnote}}
\footnotetext[1]{Angle relative to that toward Sgr A* in the GC.}
\footnotetext[2]{Reference for distance: (1) - \citet{Gaia_colab_2021}, (2) - \citet{Negueruela_2022}, (3) - \citet{Clark_2018}, (4) - \citet{Clark_2021}, (5) - \citet{Bailer_Jones_21}, (6) - \citet{Maryeva_2022}.}

\end{minipage}



\bsp	
\label{lastpage}
\end{document}